%% file: arxiv_vrperception_main.tex
\documentclass[acmlarge, nonacm]{acmart}
\pdfoutput=1

\newcommand{\mode}{1}

\graphicspath{{figures/}{pictures/}{images/}{./}} 

\usepackage{enumitem}

\setcopyright{none}
\copyrightyear{2021}
\acmYear{2021}
\acmDOI{10.1145/1122445.1122456}
\renewcommand\footnotetextcopyrightpermission[1]{}
\acmJournal{TAP}
\acmVolume{18}
\acmNumber{2}
\acmArticle{5}
\acmMonth{5}

\usepackage[capitalize]{cleveref}
\usepackage{subcaption}
\usepackage{wrapfig}

\definecolor{darkorange}{RGB}{255,140,0}
\definecolor{darkgreen}{RGB}{10,150,10}
\definecolor{darkred}{RGB}{230,12,0}
\definecolor{darkviolet}{RGB}{126,0,230}
\definecolor{darkmagenta}{RGB}{230,0,219}
\definecolor{darkyellow}{RGB}{195,205,0}

\definecolor{participant}{RGB}{90,90,90}

\definecolor{matlabgreen}{RGB}{51,160,44}
\definecolor{matlabblue}{RGB}{31,120,180}
\definecolor{matlabred}{RGB}{227,26,28}
\definecolor{matlaborange}{RGB}{255,127,0}

\newcommand{\CU}{\mathcal{U}}

\newcommand{\pS}{\mathcal{S}}
\newcommand{\pR}{\mathcal{R}}

\newcommand{\refer}{\mathrm{ref}}
\newcommand{\comp}{\mathrm{comp}}

\newcommand{\pquote}[1]{\textcolor{participant}{{``#1''}}}

\DeclareMathOperator*{\argmin}{arg\,min}

\setlist[itemize]{label=---, itemsep=2pt, topsep=2pt,leftmargin=3.5ex}

\newcommand{\approxcustom}{\raise.17ex\hbox{$\scriptstyle\sim$}}

\newlength{\halfwidth}
\begin{document}

\title{Thinking Outside the Lab: VR Size \& Depth Perception in the Wild}

\author{Rahul Arora}
\affiliation{%
  \institution{University of Toronto}
  \streetaddress{40 St. George Street}
  \city{Toronto}
  \state{Ontario}
  \country{Canada}
  }
\orcid{0000-0001-7281-8117}
\email{arorar@dgp.toronto.edu}

\author{Jiannan Li}
\affiliation{%
  \institution{University of Toronto}
  \streetaddress{40 St. George Street}
  \city{Toronto}
  \state{Ontario}
  \country{Canada}
  }
\email{jiannanli@dgp.toronto.edu}

\author{Gongyi Shi}
\affiliation{%
  \institution{University of Toronto}
  \streetaddress{40 St. George Street}
  \city{Toronto}
  \state{Ontario}
  \country{Canada}
  }
\email{gongyi.shi@mail.utoronto.ca}

\author{Karan Singh}
\affiliation{%
  \institution{University of Toronto}
  \streetaddress{40 St. George Street}
  \city{Toronto}
  \state{Ontario}
  \country{Canada}
  }
\email{karan@dgp.toronto.edu} 

\renewcommand{\shortauthors}{Arora et al.}

\begin{abstract}
Size and distance perception in Virtual Reality (VR) have been widely studied, albeit in a controlled laboratory setting with a small number of participants.
We describe a fully remote perceptual study with a gamified protocol to encourage participant engagement, which allowed us to quickly collect high-quality data from a large, diverse participant pool (N=60).
Our study aims to understand medium-field size and egocentric distance perception in real-world usage of consumer VR devices. We utilized two perceptual matching tasks---distance bisection and size matching---at the same target distances of 1--9 metres.
While the bisection protocol indicated a near-universal trend of nonlinear distance compression, the size matching estimates were more equivocal.
Varying eye-height from the floor plane showed no significant effect on the judgements.
We also discuss the pros and cons of a fully remote perceptual study in VR, the impact of hardware variation, and measures needed to ensure high-quality data.
\end{abstract}

\keywords{Depth perception, size perception, virtual reality, remote studies, gamification.}

\ccsdesc[500]{Human-centered computing~Virtual reality}
\ccsdesc[500]{Computing methodologies~Perception}

\begin{teaserfigure}
  \centering
  \includegraphics[width=\linewidth]{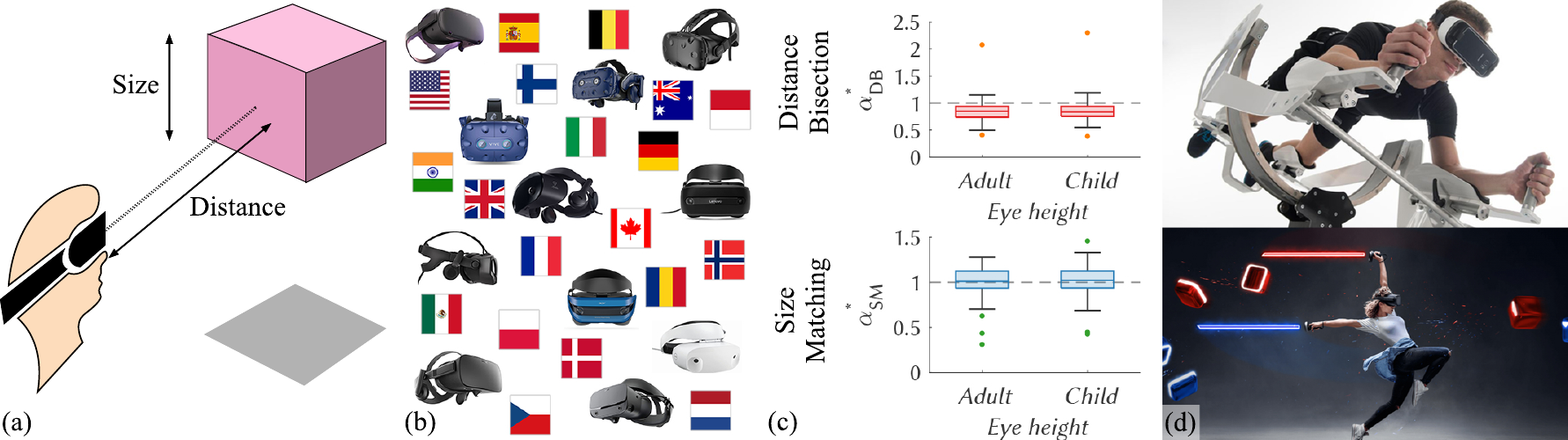}
  \caption{We conducted a remote unsupervised ``in-the-wild'' study to understand size and egocentric distance perception in VR using a gamified methodology (a). Our 60-participant study spanned 19 countries and 11 different HMDs (b). The summary of the two perceptual matching tasks is shown here (c): distance bisection judgements show a trend of nonlinear distance foreshortening (top) while size matching judgements suggest neither foreshortening nor anti-foreshortening (bottom). These measurements can inform ``distance correction'' functions that help improve user performance in applications such as VR flight simulators and games such as Beat Saber (d). \textcopyright~ICAROS (top) and Beat Games (bottom).}
  \label{fig:teaser}
\end{teaserfigure}

\maketitle

\section{Introduction}

\input{1_introduction}

\section{Related Work}
\label{sec:related}
\input{2_related}

\section{Size and Depth Perception Study}
\label{sec:study}

\input{3_study}

\section{Gamification Methodology}
\label{sec:gamification}
\input{4_gamification}

\section{Results}
\label{sec:results}
\input{5_results}

\section{Discussion}
\label{sec:discussion}
\input{6_discussion}

\section{Conclusion and Future Work}
\label{sec:conclusion}
\input{7_conclusion}

\bibliographystyle{abbrv-doi}

\bibliography{references}

\end{document}


\title{Thinking Outside the Lab: VR Size \& Depth Perception in the Wild Supplemental Document S1}

\author{Rahul Arora}
\affiliation{%
  \institution{University of Toronto}
  \streetaddress{40 St. George Street}
  \city{Toronto}
  \state{Ontario}
  \country{Canada}
  }
\orcid{0000-0001-7281-8117}
\email{arorar@dgp.toronto.edu}

\author{Jiannan Li}
\affiliation{%
  \institution{University of Toronto}
  \streetaddress{40 St. George Street}
  \city{Toronto}
  \state{Ontario}
  \country{Canada}
  }
\email{jiannanli@dgp.toronto.edu}

\author{Gongyi Shi}
\affiliation{%
  \institution{University of Toronto}
  \streetaddress{40 St. George Street}
  \city{Toronto}
  \state{Ontario}
  \country{Canada}
  }
\email{gongyi.shi@mail.utoronto.ca}

\author{Karan Singh}
\affiliation{%
  \institution{University of Toronto}
  \streetaddress{40 St. George Street}
  \city{Toronto}
  \state{Ontario}
  \country{Canada}
  }
\email{karan@dgp.toronto.edu} 

\renewcommand{\shortauthors}{Arora et al.}



\maketitle

\section{Experimental Procedure Details}

In this section, we provide additional details on participant demographics and gamification procedure.

\subsection{Full Demographics Details}
\label{sec:demographics}

\begin{wrapfigure}[16]{r}{7cm}
    \centering
    \vspace{-2.75em}
    \includegraphics[width=\linewidth]{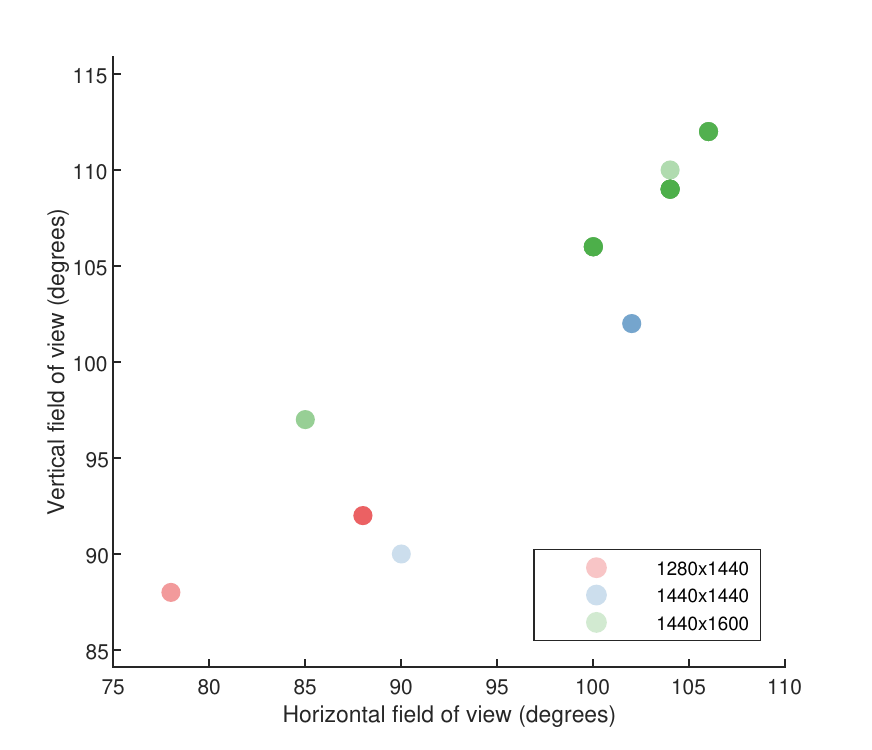}
    \caption{Field of view reported by all the headsets utilized in the study, coloured by the marketed screen resolution per eye.}
    \label{fig:hmd}
\end{wrapfigure}
Participants' countries of residence are shown in \cref{tbl:country}, while VR device usage statistics are displayed in \cref{tbl:device}. The inset figure shows the fields of view and resolutions of the headsets utilized.
We had also asked two questions to judge participants' experience with 3D distance and size judgement tasks---experience with 3D action games such as first-person shooters and with 3D design and modelling tools---on 5-point Likert scales, with 5 being the highest degree of self-reported experience. Participants were highly experienced with 3D games, with 38 answering 5 and 15 answering 4 (median 5). Experience with 3D design and modelling tools was lower (median 2), with only 5 participants reporting extensive experience with such software.

\begin{table}
\begin{minipage}[t]{.38\textwidth}
    \centering
    \captionof{table}{Participants' countries of residence.}
	\label{tbl:country}
    \rowcolors{2}{gray!25}{gray!10}
    \begin{tabular}{lr}
        \toprule
        \textbf{Country} & \textbf{\# participants} \\
        \midrule
        United States & 36 \\
        Canada & 5 \\
        France & 2 \\
        Germany & 2 \\
        Belgium & 1 \\
        Czech Republic & 1 \\
        Denmark & 1 \\
        Finland & 1 \\
        India & 1 \\
        Indonesia & 1 \\
        Italy & 1 \\
        Mexico & 1 \\
        Netherlands Antilles & 1 \\
        Norway & 1 \\
        Poland & 1 \\
        Romania & 1 \\
        Spain & 1 \\
        United Kingdom & 1 \\
        \bottomrule
    \end{tabular}
\end{minipage}\hfill
\begin{minipage}[t]{.5\textwidth}
    \centering
    \captionof{table}{Full list of devices used by the participants.}
	\label{tbl:device}
    \rowcolors{2}{gray!25}{gray!10}
    \begin{tabular}{lr}
        \toprule
        \textbf{Device} & \textbf{\# participants} \\
        \midrule
        Valve Index & 17 \\
        Oculus Quest & 15 \\
        HTC Vive & 10 \\
        Oculus Rift S & 5 \\
        Lenovo Explorer (Windows MR) & 3 \\
        Samsung Odyssey+ (Windows MR) & 3 \\
        Oculus Rift & 2 \\
        HTC Vive Pro & 2 \\
        HTC Vive Pro Eye & 1 \\
        Acer AR101 (Windows MR) & 1 \\
        Dell Visor (Windows MR) & 1 \\
        \bottomrule
    \end{tabular}
		\vspace{2em}
		
		\includegraphics[width=.75\linewidth]{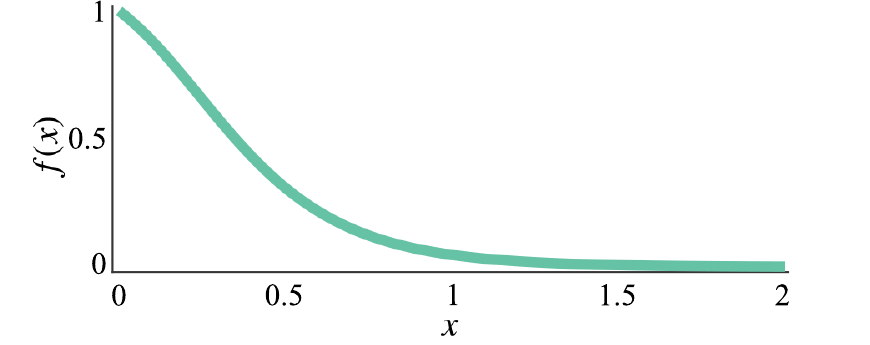}
		\captionof{figure}{Scoring function.}
		\label{fig:scoring}
\end{minipage}
\end{table}

\subsection{User Controls Details}
\label{sec:controls}

The study controls were designed to be simple and easily extensible to diverse controllers. Participants could use any available one-dimensional input button to control the size of the \emph{comparison} cube in the \emph{size matching} tasks. This was chosen to be the y-axis of a thumbstick or trackpad on the controller. As participants held it near one of the extremes for time $t$, the size $S_{comp}$ was increased (or decreased) by an amount $c \mathrm{v}(t) \Delta t$, where $\Delta t$ is the time since the last update, $c\in\mathbb{R}^+$ is a positive constant, and $\mathrm{v}(t)$ is the rate of change at time $t$. In order to reduce the impact of the kinetic depth information and to allow large changes to the cube size efficiently, we model $\mathrm{v}(t)$ with an acceleration term. That is, $\mathrm{v}(t)=at$, where $a\in\mathbb{R}^+$ is a positive constant. The \emph{comparison} cube distance is controlled analogously in the \emph{distance bisection} tasks.

The only other control is a ``confirm'' button, for which we use the primary push button on the controller, typically called the ``A'', ``X'', or ``Menu'' button, depending on the device.

\subsection{Gamification Scoring Details}
\label{sec:gamification}

\begin{wrapfigure}[24]{r}{\halfwidth}
    \centering
	\vspace{-1em}
    \includegraphics[width=\linewidth]{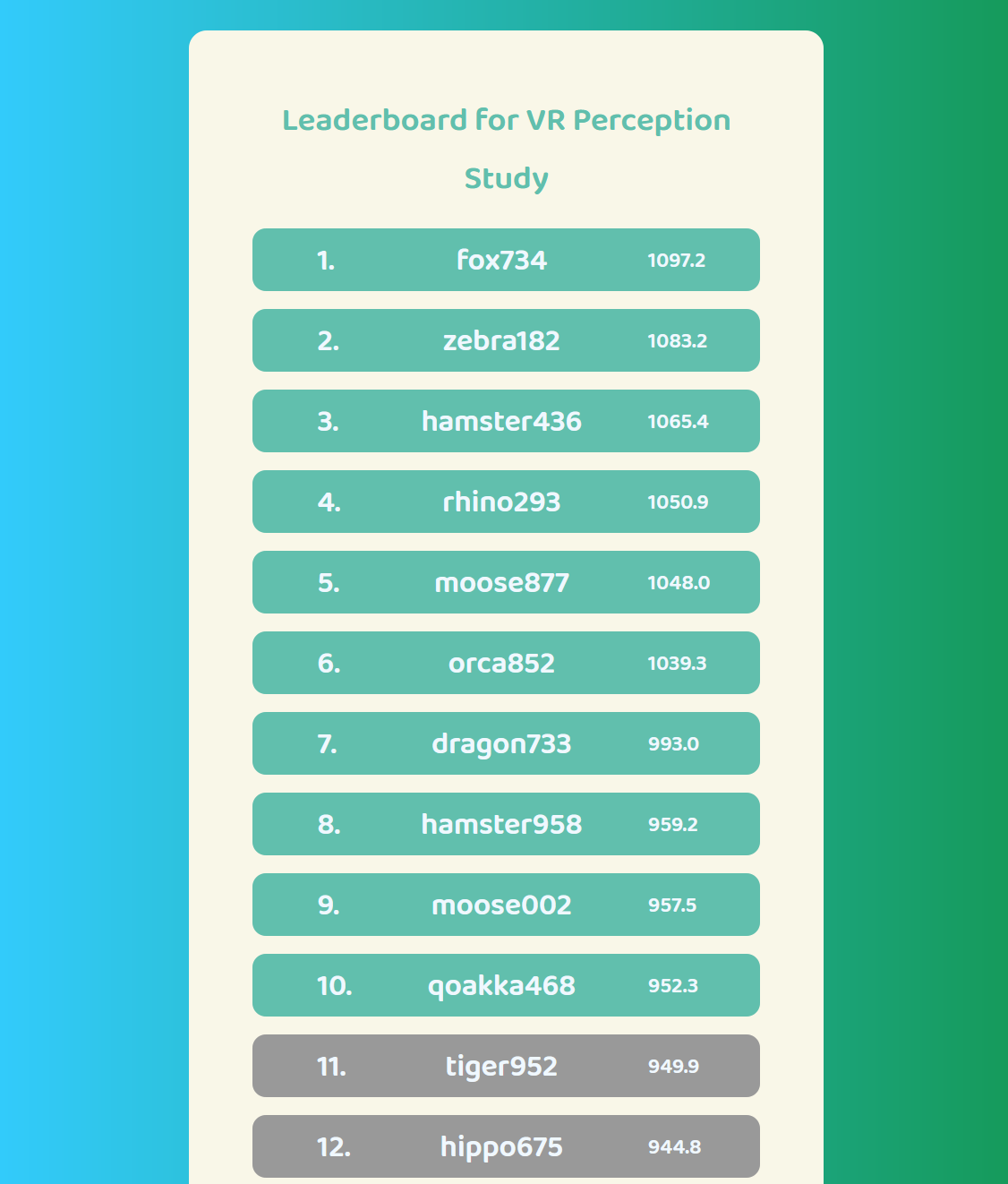}
    \caption{Screenshot of the online leaderboard (vertically clipped).}
    \label{fig:leaderboard}
\end{wrapfigure}

Participants had access to an online leaderboard to see their score, as well as their position relative to others. The top ten participants received an additional gift card worth US \$15. The leaderboard is shown in \cref{fig:leaderboard}.

The score was computed out of 1000, along with an additional bonus of 100 points each for the second and third session, leading to a maximum possible score of 1200. The score for each experiment block $\Xi_{block}$, out of a total of 250, was computed as follows.

\begin{equation}
    \Xi_{block} = \Xi_{base} + \xi_{trial} \left( \sum_{t\in \{1,\ldots, 15\}} \mathrm{f}_t \right)
\end{equation}

The base score $\Xi_{base}$ was set to 25 points, thus participants received at least 100 points for a session, irrespective of their responses. The per-trial maximum score $\xi_{trial}$ was 15. $\mathrm{f}_t$ gives the score for each of the 15 trials in a block.

\begin{equation}
    \mathrm{f}_t = f(x) = \left( 1 - \frac{x}{x+\exp(-3x)}\right).
\end{equation}

This sigmoid-like function gives a maximum when the difference between the accurate response and the chosen response is 0, and increases symmetrically when it deviates to either direction (\cref{fig:scoring}). Here $x$ is a transformation applied on the participant's response as follows. In a \emph{size matching} trial, let the size difference corresponding to the participant's response be $y = S_{comp} - s_{ref}$. Then we set $x$ to be $|y| / 0.3$. Similarly, for the \emph{distance bisection} trials, given the chosen \emph{comparison} distance $R_{comp}$, we set $x = |2R_{comp} - r_{ref}| / 0.3$.

\subsection{Study Deployment and Implementation Details}
\label{Sec:deployment}

Please note that all material related to study deployment is included in supplemental ZIP archive \textbf{S3}. Participation calls on Facebook and Reddit linked participants to a static webpage (\textbf{S3d}) that provided essential study and compensation details. A Google Form\footnote{\url{https://docs.google.com/forms}} (\textbf{S3e}) accessible from the webpage first obtained participant consent, followed by a second screen showing the short (2.5 minute) instructions video (\textbf{S3b}) hosted on YouTube\footnote{\url{https://www.youtube.com}}. While Google Forms does not provide a method to ensure that participants watched the instructions video till the end, YouTube statistics confirmed that most participants watched the bulk of the video. As an additional safeguard to ensure that participants understood the tasks, the study procedure included explanatory text and practice trials for both the \emph{protocols}. In the third section of the Form, participants were provided with application download links (\textbf{S3g--i}). The study application was developed in the C\# programming language using the Unity engine\footnote{\url{https://www.unity.com}}.

After completing a session, the application supplied a randomly-generated username to the participant, which was used to authenticate the remaining survey portion of the Form. Survey questions were aimed at understanding participants' strategies and the perceived differences between various conditions.

Participants were not required to complete any of the above steps for additional sessions. They could simply restart the downloaded program and complete another session. A unique ID associated with the participant's device ensured that they could not complete more than three sessions. All the study data was uploaded to a Back4App server\footnote{\url{https://www.back4app.com}} hosting the open source Parse platform\footnote{\url{https://parseplatform.org/}}.







\section{Results with First Session Data Only}
\label{sec:firstSession}

In the main document, we reported results for mean size ratio for \emph{size matching} tasks and mean distance ratio for \emph{distance bisection} tasks and the respective standard deviations. The reported results utilize participant judgements across all the sessions (recall that many participants completed multiple sessions). The mean size ratio was $1.094\pm0.428$, while the mean distance ratio was $0.469\pm0.157$. If we only utilize data from participants' first session, thus weighing all participants equally for computing summary statistics, the mean size ratio was found to be $1.091\pm0.464$ and the mean distance ratio $0.464\pm0.178$. Clearly, the mean values are very similar, consistent with our observation that no learning effects were observed.






%% file: 1_introduction.tex
Understanding human perception in Virtual Reality (VR) is a fundamental question in VR research. From casual applications such as games and movies to performance-critical technical training, modelling biases in people's perception of their surroundings in VR is an important endeavour. Unsurprisingly then, ever since the early days of VR, visual perception has received considerable attention in psychology and cognitive science research~\cite{beall1995absolute,bingham2001accommodation, eggleston1996virtual}. Of particular interest is the perception of the geometric attributes of visual space, namely distance and size. Real-world size and distance perception experiments have a rich century-old history in cognitive science~\cite{cutting2003reconceiving,wagner2012sensory,howard2012perceiving}, and similar experiments in VR serve a dual purpose. Not only do such studies advance our understanding of visual perception in VR, but a virtual world closely simulating reality provides a unique opportunity to construct environments that are difficult or impossible in the real world. Such fantastical constructs can be used to manipulate perceptual cues at will and gain insights into the human visual system~\cite{bruder2011tuning}. VR size and distance perception studies can thus improve our understanding of real-world perception as well. At the same time, due to the shortcomings of current VR devices, such as the well-known vergence-accommodation conflict~\cite{hoffman2008vergence}, it is essential to conduct perceptual studies in VR to inform the design of VR applications where perceptual accuracy is crucial.

Most geometric perception studies, whether real-world or in VR, utilize strictly controlled laboratory conditions (e.g. \cite{gogel1987familiar,eby1995perceptual,glennerster2006humans,peillard2019virtual}). While maintaining strict control over the experiment is invaluable for gaining theoretical insights into the human visual system as noted above, the practical utility of the results to real-world applications is limited. For example, a typical perceptual experiment in VR would utilize the same hardware for all users, ensure strict calibration of the interpupillary distance (IPD), and control the position of the head-mounted display (HMD) on the user's head. In real-world usage, users utilize diverse VR setups, and are typically untrained in proper IPD adjustment---adjusting IPD and HMD position for comfort, rather than physical accuracy. Recently, Hornsey et al.~\shortcite{hornsey2020size} reported encouraging results on a size and shape constancy task where a) two different devices, Oculus Rift and HTC Vive, were utilized and b) IPD was not adjusted or controlled for. However, the results were still conducted in a laboratory setting, requiring significant investment of experimenter and participant time and effort. Inspired by their results, we present the first remote study on geometric perception in VR (\cref{fig:teaser}a). Our size and distance perception experiment aims at achieving results that are \emph{ecologically valid}, that is, we want to model real-world HMD usage to inform the development of VR applications. In addition to removing strict controls on the experiment conditions, a remote study enables us to access a large and diverse pool of participants (\cref{fig:teaser}b). In this work, we report on a medium-scale study with N=60. Studies such as ours are especially relevant today owing to the COVID-19 physical distancing requirements which make remote studies an attractive option for experimental research involving human subjects.

\paragraph{Study Summary.}
\label{sec:summary}

In our experiment, we study the perception of size and egocentric distance in VR using two different measurement protocols. The first protocol involves a classical size matching task~\cite{holway1941determinants,smith1953methodological}, where participants resize a nearby \emph{comparison} object to match the physical size of a relatively distant \emph{reference} object. Judgements of perceived egocentric distance can be achieved by assuming the size-distance invariance hypothesis (SDIH)~\cite{epstein1963attitudes} (also see Howard~\cite[Sec. 29.3.2]{howard2012perceiving}, Wagner~\cite{wagner2012sensory}) and inverting the perceived size judgements. The classical SDIH posits that given a visual (retinal) size, the perceived size of an object is proportional to the perceived distance to it.
Our second protocol---distance bisection---elicits distance perception judgements directly:  participants move the \emph{comparison} object so that it lies half as far from them as a fixed \emph{reference} object.
Thus, the \emph{reference} object is placed at similar positions for both the protocols, while the \emph{comparison} is farther in the distance bisection protocol as compared to the size matching protocol. 
While our protocol choices are similar to Gruber~\cite{gruber1954relation}, who applied both protocols and compared their results to evaluate SDIH in real environments, we do not formally evaluate the applicability of SDIH in our scenario due to the differences in the egocentric position of the \emph{comparison}.
In addition, our protocols elicit perceived distance \emph{relative} to the physical distance, but do not help estimate the perceived \emph{absolute} distance.
This relative measure indicates whether the visual space exhibits \emph{foreshortening}, that is, fixed depth intervals being increasingly compressed with further distance from the observer, or \emph{anti-foreshortening}, where the visual space is expanded. Accurate judgements indicate that the perceived distance varies linearly with the physical distance and the visual space is \emph{invariant}.

A second factor, explored with both the measurement protocols, is eye height. We compare distance and size perception when the scene was rendered assuming an eye height similar to an average adult (170 cm) vs. a much shorter height of 50 cm. This investigation is motivated by two disparate applications---seated or short characters (child, dwarf, or quadruped) in VR games, and the use of VR for tele-operating robots that are often short~\cite{corujeira2013stereocopic}.

A common challenge with remote studies is control over data quality. Malicious participants can develop strategies to game the system, and the collected data may need thorough analysis to detect cheating~\cite{snow2008cheap,hirth2011cost,galais2014you}. An additional challenge is the lack of participant engagement~\cite{galais2014you}, which can reduce data validity as the experiment goes on. We designed a gamified experiment to tackle these challenges: participants were awarded points for their perceptual judgements, with more accurate judgements getting a higher score. An online leaderboard prompted participants of their rank vis-\`{a}-vis others, and top participants received a bonus compensation. Further, we broke down the tasks into easily-fulfilled ``chunks'' to prevent VR cybersickness or boredom due to the monotonous tasks, thereby ensuring high quality of the collected data.

\paragraph{Contributions.}
\label{sec:contribution}

Following are our main contributions.

\begin{itemize}
    \item The first fully remote study on geometric perception in VR.
    \item Understanding how perceived distance $\pR$ varies with actual egocentric distance $R$ using two different perceptual matching protocols. With size matching, the overall trend did not point clearly to either non-linear foreshortening or anti-foreshortening: $\pR = kR^{1.001}$. In contrast, with distance bisection, a clear trend of foreshortening emerged: $\pR = kR^{0.863}$.
    \item Testing size and depth perception with two different eye heights. Perceptual judgements were similar across the two eye heights.
    \item A gamified protocol for our remote study and a discussion on adapting it for other perceptual studies in VR.
\end{itemize}



%% file: 2_related.tex
Our work straddles  three themes: understanding biases in the size and depth perception, modelling human perception in virtual environments, and remote experiments for VR and perceptual studies.


\subsection{Bias in Human Depth and Size Perception}
\label{sec:bias}

Fundamental to human vision, size and depth perception have received sustained interest in psychology and cognitive science. Howard's book~\shortcite{howard2012perceiving} provides a comprehensive review of this broad area.
Size and depth perception biases reveal basic mechanisms of the vision system~\cite{loomis1992visual} and are critically relevant to efficiency and safety in tasks such as aircraft piloting~\cite{lappin2006environmnet}. 
Prior research has presented rich and sometimes contrasting results about biases in egocentric depth perception.
While some studies found that people can perceive depth accurately and reliably~\cite{rieser1990visual,loomis1996visual,purdy1955distance}, others showed that they are prone to systematic errors~\cite{lappin2006environmnet,foley2004visual,gilinsky1951perceived}.
A large body of research noted the effect of foreshortening \cite{gilinsky1951perceived, toye1986effect, wagner1985metric, norman1996visual, loomis1992visual}. The effect was found to be stronger with depth cues further removed~\cite{philbeck1997comparison}. However, several studies using a bisection protocol presented the opposite evidence in support of an invariant or expanded visual space. Rieser et al.~\shortcite{rieser1990visual} reported that participants were largely accurate in judging the mid-points of 4--24 metre self-to-target distances in an open field. Lappin et al.~\shortcite{lappin2006environmnet} found an environment-sensitive antiforeshortening effect; participants perceived the mid-points of 15 and 30 metre distances as further away than the true mid-points. These contradictory results remain unexplained yet.

Size perception is closely related to depth perception and the two are commonly measured together in prior studies~\cite{gogel1987familiar,haber2001independence}.
The size-distance invariance hypothesis (SDIH)~\cite{epstein1963attitudes,epstein1961current} directly links the two, positing that with a fixed visual angle, the perceived size is proportional to the perceived distance. Researchers have applied SDIH to measure perceived distance through perceived size as a proxy~\cite{loomis2003visual,kelly2017perceived}. However, the validity of SDIH is controversial~\cite{epstein1961current}. 
Through size-matching tasks, a number of experiments~\cite{smith1953methodological,jenkin1957effects,jenkin1959relationship} have found that the sizes of further placed objects are increasingly overestimated, inconsistent with the commonly observed foreshortening effect if SDIH holds true. More directly, Gruber~\cite{gruber1954relation} showed that participants underestimated the sizes of objects in size matching but overestimated the distances to the same objects in distance bisection. Various attempts have been made to resolve the size-distance paradox~\cite{ross2003levels}. Some researchers argue that size and depth perception are indeed independent processes~\cite{haber2001independence,day1989exorcize}.


Our study revisits biases in these two important aspects of spatial perception in an immersive virtual environment, using a size matching and a distance bisection protocol employed in classical work (e.g. Holway and Boring~\shortcite{holway1941determinants} and Gruber~\cite{gruber1954relation}). The virtuality of the computer-generated world allows for conducting studies at a much larger scale (N=60) in comparison to traditional perception studies which typically involved fewer than 20 participants. While the virtual and physical worlds are perceived notably differently, we hope that data collected from large and diverse sample pools can also contribute to deepening our understanding of human vision.

\subsection{Perception Studies in Immersive VR}
\label{sec:virtualStudies}

Presenting a virtual space that can be accurately perceived by human eyes is important for VR applications to be more immersive and useful. 
In contrast to the somewhat contradictory results about depth and size perception in the real world, research on egocentric depth perception in VR has, in general, found a tendency of depth underestimation~\cite{renner2013perception} and size underconstancy~\cite{hornsey2020size,kelly2017perceived}. 
In a study replicating the real-world experiment of Lappin et al.~\shortcite{lappin2006environmnet} in a virtual setting~\cite{bodenheimer2007distance}, participants placed the perceived mid-points closer than the true mid-points, in contrast to the antiforeshortening effect observed by Lappin et al. in the real-world setting.
Corujeira and Oakley~\shortcite{corujeira2013stereocopic} also found depth underestimation in VR using a blind walking protocol. They further pointed out the effect of eye height on depth perception accuracy; a lying-on-the-floor eye height of 20 cm led to more accurate judgements than when seated (110 cm). Size-matching studies in VR have found underestimation of distant object sizes~\cite{hornsey2020size, kelly2017perceived,kenyon2007size}, largely compatible with results from studies on depth perception.

In addition to the differences in the general tendency of biases from the real world, depth perception in VR has also been found to be less precise (higher depth discrimination threshold) than in the real world~\cite{naceri2015depth}. This discrepancy between real and virtual environments has been of much research interest. We refer the reader to the survey by Renner et al.~\shortcite{renner2013perception} for an in-depth review on the topic. The significant variability of available pictorial depth cues---such as texture~\cite{thomas2002surface}, lighting~\cite{tai2012daylighting}, number of objects~\cite{kenyon2007size}---in different virtual environments could cause varying levels of distortion in perceived depth. The technical properties of current VR systems, such as FoV~\cite{jones2012comparability}, IPD~\cite{willemsen2008effects,bruder2012analyzing}, and vergence-accommodation conflict~\cite{hoffman2008vergence} may also affect depth perception accuracy.

The clear influence of hardware configurations on perception in VR may suggest that only tightly controlled lab studies offer a reliable avenue for studying VR perception. However, recent work has reported encouraging results with uncalibrated consumer VR devices~\cite{hornsey2020size}, consistent with controlled studies with lengthy calibration processes. Our work further extends VR perception studies outside the laboratory to gather data from a large, diverse, sample pool and evaluate the feasibility of this promising methodology.

\subsection{Unsupervised Remote Studies and Gamification}
\label{sec:crowdStudies}

Online studies have recently gained strong traction for a variety of research topics, including decision making~\cite{mason2012conducting}, social behaviour~\cite{ma2018international,gehlbach2015many}, perception~\cite{woods2015conducting,sasaki2019crowdsourcing}, design optimization~\cite{koyama2014crowd,koyama2017sequential}, and user interface performance~\cite{komarov2013crowdsourcing}.
Researchers have noted their potential benefits of reaching the ``internet scale''~\cite{reinecke2015labinthewild,mason2012conducting}: enabling running parallel sessions to drastically reduce study turnover time and accessing large participants pools beyond the researchers' local communities.
Despite the usually unsupervised and uncontrolled setting of these studies, past research has shown that proper measures can ensure high-quality data collection~\cite{mason2012conducting,woods2015conducting}.

Remote VR studies~\cite{ragozin2020mazerunvr,huber2020conducting,ma2018international,steed2016wild,mottelson2017virtual} are quickly becoming feasible with the increasing dissemination of consumer VR devices. 
Recent efforts have shown that online VR studies can  efficiently gather reliable data~\cite{steed2016wild,mottelson2017virtual} and replicate established results~\cite{ma2018international,huber2020conducting}. Mottelson and Hornb{\ae}k~\shortcite{mottelson2017virtual} distributed VR devices to participants, who were asked to test two input techniques and a VR illusion study in an \emph{in-the-wild} setting. They found effects comparable to traditional laboratory settings , albeit with larger variation. Ma et al.~\shortcite{ma2018international} were able to successfully induce behavioural manipulation in three VR studies conducted through Amazon Mechanical Turk. Huber and Gajos~\shortcite{huber2020conducting} replicated past results obtained from conventional lab settings for a spatial navigation task and a negotiation task with uncompensated VR users.

While results from prior works are encouraging, researchers have also noted uncertainties and difficulties in remote VR data collection, including variations in hardware~\cite{ma2018international}, unknown instruction effectiveness~\cite{steed2016wild}, and ethical concerns~\cite{steed2016wild,mottelson2017virtual,snow2008cheap,hirth2011cost}. In particular, crowdsourced studies tend to face diminishing engagement from participants, who may optimize for return on effort rather than quality~\cite{gould2016diminished,hata2017glimpse,galais2014you}. Gamification has been identified as a method to sustain engagement and increase data quality~\cite{looyestyn2017does,van2017gamification}. Introducing \emph{intrinsic} motivations, such as helping participants know themselves better, can have similar positive effects on engagement~\cite{huber2020conducting}.

Experiments on low-level human perception have typically been conducted in supervised and strictly controlled environments to ensure internal validity. Our study explores the potential of an unsupervised, fully-remote protocol for collecting high-quality low-level perception data in VR. While building on recommended practices suggested by prior work, we introduce additional measures to combat data quality challenges unique to our study tasks, such as limiting head motion. We also add gamification mechanisms and intrinsic motivations to our study design to improve participant engagement.

%% file: 3_study.tex
We will first describe the perceptual experiment, before detailing our gamification strategy.

\begin{wrapfigure}[12]{r}{\halfwidth}
	\centering
	\vspace{-1em}
	\includegraphics[width=\linewidth]{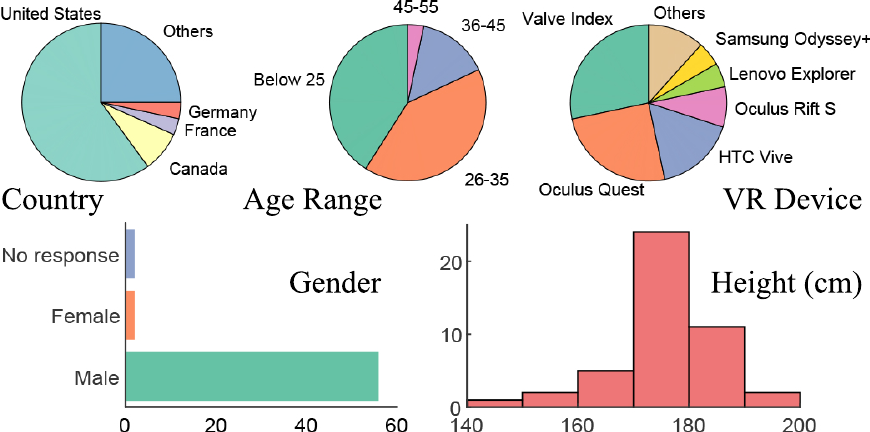}
	\caption{Demographics information of the study participants.}
	\label{fig:demographics}
\end{wrapfigure}
\subsection{Participants}
\label{sec:participants}

Sixty participants were recruited via two main channels: a) VR interest groups on social media (Facebook/Reddit) and b) HCI/VR research e-mail lists. Almost all were male (56M, 2F, 2 unspecified), reflecting current estimates of VR ownership \cite{uploadvr2017report,rakuten2019mobile}. Most were young (25: 18--25, 24: 26--35), but some were middle-aged (9:  36--45, 2: 46--55). Participants hailed from 19 countries across 5 continents; however, over half (36) were from the USA. See \cref{fig:demographics} and supplemental \textbf{S1} (Sec.~1.1) for details.
Participants received US \$5 (or equivalent), and the top ten on the gamified leaderboard (\S~\ref{sec:gamification}) were paid an additional \$15.

\subsection{Apparatus}
\label{sec:apparatus}

All participants had access to a six--degree of freedom (6-DoF) VR device with a controller. They used 11 different devices, with Valve Index (17) and Oculus Quest (15) being the most popular. Device resolutions ranged between 1280$\times$1440 and 1440$\times$1600 per eye, while the Field of View (FoV) lay between $78^\circ\times88^\circ$\ and $106^\circ\times112^\circ$.

\subsection{Stimuli}
\label{sec:stimuli}

\begin{figure}
    \centering
    \includegraphics[width=\textwidth]{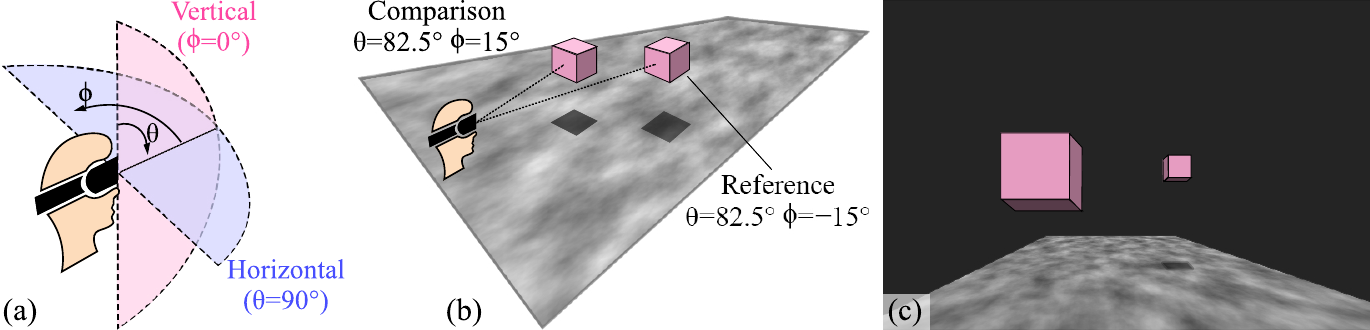}
    \caption{The spherical coordinate system used in setting up the stimuli (a). An illustration of the stimuli, showing the \emph{reference} and \emph{comparison} cubes, the Perlin-noise textured ground plane, and the cast shadows (b). The actual distance to the \emph{comparison} depends on the \emph{protocol}: for \emph{size matching}, it is positioned very close (75 cm) to the user, while for \emph{distance bisection}, it is placed farther. An illustration of the scene as seen from the subject's perspective (c).}
    \label{fig:stimuliPosition}
\end{figure}

In both the protocols, a minimal scene containing only the fixed \emph{reference} object, the subject-controlled \emph{comparison} object and a ground plane were rendered over a solid dark grey background. Both the objects were dull-pink coloured cubes and the ground plane was textured with grey Perlin noise~\cite{perlin2002improving}. Objects more complex than cubes could have been chosen, but pilot testing showed no significant difference between cubes and complex objects such as tori and Stanford bunnies. Further, the colour choice for the cubes was arbitrary, but the colour saturation was used to clearly contrast the foreground objects against the background and ground plane consisting of greys only. In addition to ambient lighting, a directional light pointing downwards lit the scene, causing the cubes to cast shadows on the plane. The shadows were meant to emphasize the difference between the two \emph{eye heights} to the participants. No additional context or environmental cues were provided (\cref{fig:stimuliPosition}b). The off-the-ground object positioning is motivated by mid-air objects common in popular VR games such as Beat Saber and Superhot, and creation applications such as Tilt Brush and Gravity Sketch.


The stimuli were positioned relative to the participant in a spherical coordinate system (\cref{fig:stimuliPosition}) centred at the participant's cyclopean eye (point midway between the eyes), with a vertical zenith ($\theta=0^\circ$) and horizontal reference plane ($\theta=90^\circ$). The initial cyclopean line of sight projected onto the horizontal served as the azimuth reference direction ($\phi=0^\circ$). Note that in this coordinate system, a point is represented by the vector $(r, \theta, \phi)$, where $r$ is the egocentric distance, or interchangeably, the depth from the participant.

In all the conditions, both the cubes were positioned near the participant's fovea. The \emph{reference} cube was positioned so that the centre of its front face was slightly to the right of the line of sight ($\phi_{\refer}=-15^\circ$) and slightly above the horizontal ($\theta_{\refer}=82.5^\circ$) and the \emph{comparison} positioned at the same inclination, but to the left ($\phi_{\comp}=15^\circ$). For different trials, the distance $r$ to the \emph{reference} cube was varied between 1 and 9 metres, with 2m increments, that is, $r_{\refer}\in\{1\text{m}, 3\text{m}, 5\text{m}, 7\text{m}, 9\text{m}\}$. Note again, that similar to the $\phi$ and $\theta$ coordinates, the distance is to the centre of the front face (the one facing the subject) of the cube. The \emph{reference} cube's size $s_{\refer}$ was chosen uniformly randomly between 25 and 35 cm; the randomization intended to reduce the chance of inadvertently turning the \emph{reference} cube into a familiar-sized object.

Task-relevant distance cues thus available to the participants were: stereopsis, convergence, motion parallax, perspective, lighting \& shading, shadows, and to a degree, texture gradient on the ground plane. We now describe protocol-dependent stimuli characteristics.

\paragraph{Distance Bisection Protocol.}
For each trial, the distance $r_{\comp}$ as well as size $s_{\comp}$ of the \emph{comparison} were chosen so that the position was not close to the accurate response ($r_{\comp} \not\approx r_{\refer}/2$), and the size was not close to that of the \emph{reference}: $s_{\comp}\not\approx s_{\refer}$. These choices were made to avoid participants simply choosing the initial position as their response and to avoid the use of size matching as an effective cue, respectively. Specifically, for setting $r_{\comp}$, we first chose uniformly randomly whether to make it larger (75\% of $r_{\refer}$) or smaller (25\% of $r_{\refer}$) than the correct response, and then added a uniform random perturbation of 10\%. Thus, 

\vspace{-1em}
\begin{equation}
    r_{\comp} = (f \pm \epsilon) r_{\refer},\ \text{where}\ \ f\sim\CU\{0.25, 0.75\},\ \epsilon\sim\CU(0, 0.1)\text{.}
    \label{eq:rcomp}
\end{equation}

$s_{\comp}$ was similarly chosen w.r.t. $s_{\refer}$ but with $f\sim\CU\{0.7, 1.3\}$.

\paragraph{Size Matching Protocol.}
The size and distance parameters of the \emph{comparison} were similarly initialized, with the only exception being the range of the random perturbation $\epsilon$ for $s_{\comp}$, which was sampled from $\CU(0, 0.15)$. This was motivated by the higher inter-participant dissimilarity observed in our \emph{size matching} pilots.

\subsection{Tasks and Controls}
\label{sec:task}

For the \emph{bisection task}, participants were able to control the distance of the \emph{comparison} cube by using the thumbstick on their controller. Note that $\theta_{\comp}$ and $\phi_{\comp}$ remained fixed. Participants were instructed to position the \emph{comparison} so that the center of its front face was half the distance from them as that of the \emph{reference}.

For the \emph{size matching} task, participants used the thumbstick to increase/decrease the size of the \emph{comparison}. The instruction was to match the physical size, that is, the 3D size, of the \emph{comparison} to the \emph{reference}. Note that the precise instruction is important here, as the size matching instruction can substantially impact the participants' responses (see Wagner~\cite[p.64]{wagner2012sensory}). In Wagner's terms, we used \emph{objective instructions} for the task. Readers are encouraged to scrutinize the instructions video in supplemental archive \textbf{S3} for details.

In either protocol, participants had 20 seconds to complete each trial. Participants pressed a button on their controller to confirm their response. The remaining time was always displayed, and if the response was not confirmed in the given time limit, the trial was rejected and excluded from the analysis. Pilot tests without time restrictions showed that \approxcustom10--12 s was enough to finish a trial and we therefore chose a conservative time limit of 20 s per trial.

\subsection{Procedure}
\label{sec:procedure}

The experiment utilized a 2$\times$2$\times$5 full-factorial within-subject design, with 3 repetitions for each factor combination.
For an experiment session, the order of the \emph{protocols} was randomized. Within each protocol, the order of the \emph{eye height} variable was randomized between \emph{adult} and \emph{child}. The ground plane was rendered 170 cm below eye-height for the former, and 50 cm for the latter. Trials for a (\emph{protocol}, \emph{eye height}) pair composed an experiment ``block''. Each block consisted of 15 trials in a random order: the five possible \emph{reference distances}---1m, 3m, 5m, 7m, 9m---repeated thrice each. Thus, each experiment session contained 60 trials.

For a short delay of 0.5 s between trials, no stimuli was displayed.
Before each block, a text prompt informed the participants of the upcoming condition. The first block for both \emph{protocols} started with 2 additional untimed practice trials.
An experiment session could take up to 22 mins in theory, but none took over 15 mins in practice.

\paragraph{Freedom of Movement.} As soon as participants pressed a button to start the first block, the position and orientation of their head was stored and used to construct the coordinate system described in \S~\ref{sec:stimuli}. During the experiment session, participants could move their head up to 15 cm from this initial position and rotate it up to 30$^\circ$\ from the initial orientation. Upon exceeding either limit, the stimuli were hidden and a warning message was displayed, until the participants went back to the allowed range of motion.

\subsection{Hypotheses}
\label{sec:hypotheses}

Based on existing work and pilot tests, we formulated four hypotheses. Prior work has indicated that distance to the bisection point is typically underestimated in VR~\cite{bodenheimer2007distance}, and we expected the same result (\textbf{H1}). For the \emph{size matching} task, our cue-limited situation suggests a mix of objective size matching (perfect matching) and visual size matching leading to size underconstancy~\cite{eggleston1996virtual}. Thus, we expected slight size underconstancy (\textbf{H2}).
Finally, we expected participants to be more accurate for both \emph{size matching} and \emph{distance bisection} in the \emph{adult eye height} condition (\textbf{H3}). Note that this is in contrast to existing work, where subjects were typically more accurate in judging distance when positioned close to the ground~\cite{corujeira2013stereocopic}. However, unlike most existing work, we position objects off-the-ground. Our hypothesis \textbf{H3} is mainly driven by the expectation that participants will be more accurate in the \emph{adult} condition, which is closer to their real-world experience than \emph{child eye height}.







%% file: 4_gamification.tex
Realizing that the experimental tasks for gauging size and depth perception can be rather dry and unengaging, we decided to use game elements in our study. Gamification has been shown to be an effective tool for enhancing user engagement~\cite{looyestyn2017does} and improving task performance in quantitative studies~\cite{van2017gamification}. It can also help maintain high data-quality in a remote study such as ours~\cite{looyestyn2017does}.

For each trial, participants were awarded scores for their responses, using a strictly concave function over the domain of response (\emph{comparison} size in size matching and distance in bisection), with the maximum at the accurate response (\textbf{S1}, Sec.~1.3). Participants' scores over all successful trials in each experiment block were tallied up and displayed at the very end of the session. Note that participants were only shown the aggregate scores for each block, so no feedback was given on individual trials or on whether over- or underestimation happened. Participants were encouraged to complete multiple experiment sessions, up to three. These additional sessions allowed participants to improve their score, and also get bonus points for simply completing the 2\textsuperscript{nd} and 3\textsuperscript{rd} sessions. Thus, instead of 3 judgements for each combination of experiment factors (\S~\ref{sec:procedure}), we could get up to 9 judgements for each participant. Without appropriate safeguards, repeated sessions could lead to biased judgements as participants could try to reverse-engineer the score. Our methodology eliminated this possibility by a) only providing aggregated feedback in the form of total score for each block of 15 trials and b) limiting each participant to three sessions. Thus, participants' ability to learn the mapping from the score to their responses was severely hindered and no learning effects were expected.

An online leaderboard was maintained where participants could see their (and others') current scores. To keep participants engaged with the study tasks, receive honest judgements, and to encourage the completion of additional experiment sessions, the top ten participants on the leaderboard were provided with an additional US \$15 on top of the base compensation of US \$5 for participation. Further, advertisements for the study highlighted that it allowed participants to test their size and distance perception in VR and to compete against strangers in perception tasks, thus adding intrinsic motivation as well as a competitive spin to the task.

\paragraph{Study Flow.}
Participants started by providing their informed consent. They then watched the short instructions video (2.5 min), and downloaded an executable. After starting the application and completing an experiment session, participants filled a demographics survey and provided an introspective report of the cues they utilized for the tasks. To complete additional sessions, participants could simply restart the application. We maintained participant privacy by a) assigning a randomly-generated playful username which was then displayed on the leaderboard, and b) using a hash of unique IDs associated with their device as an authentication mechanism, so no personal information was required (details in \textbf{S1}, Sec.~1.4). Participants only provided an email address for gift card delivery. Finally, participants could opt-out of the leaderboard; only one did.

\paragraph{Success of the Methodology.}
The gamification methodology was an immediate success. The study was able to reach our budgeted target of 60 participants within 5 days. A number of participants completed multiple sessions (21: $3$, 10: $2$, 29: $1$, mean: $1.9$), contributing \textbf{a total of 113 experimental sessions encompassing 3360 judgements each for size matching and for distance bisection}. Participants commented that they did the study purely out of interest in VR and/or scientific research; for example, they participated to help in \pquote{building a foundation of understanding for future technologies}, because they thought it was \pquote{definitely an interesting thing to study}, or that it was \pquote{more about the research than the money}. However, the opportunity to earn additional money was appreciated by some as \pquote{a nice bonus}.

%% file: 5_results.tex
\begin{figure*}[tb]
    \centering
    \includegraphics[width=\linewidth]{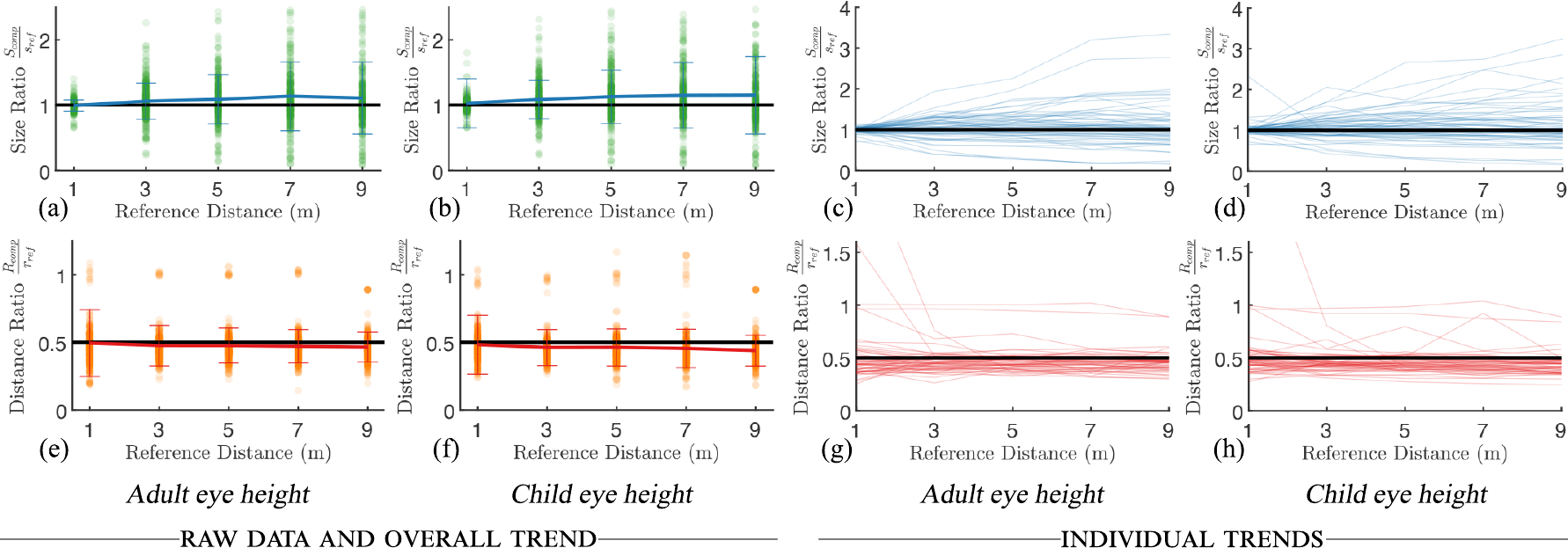}
    \caption{Results for the \emph{size matching} (top) and \emph{distance bisection} (bottom). Left and center-left columns: all recorded judgements along with the mean and standard deviation over all participants for \emph{adult} and \emph{child} height respectively. Center-right and right: mean trends for individual participants. Accurate judgements are shown in \textbf{black}, while mean judgements are shown in \textbf{\textcolor{matlabblue}{blue}}/\textbf{\textcolor{matlabred}{red}}. Notice that the distance to the bisection point is underestimated near-consistently, but for size matching, trends of both underconstancy and overconstancy are common.}
    \label{fig:allJudgements}
\end{figure*}

Raw data for all participants has been plotted in \cref{fig:allJudgements}a--b for \emph{size matching} and for \emph{distance bisection} tasks in \cref{fig:allJudgements}e--f.
Note that we denote the participants' responses in uppercase: \emph{comparison} size in \emph{size matching} with $S_{\comp}$ and \emph{comparison} distance in \emph{distance bisection} with $R_{\comp}$, to contrast with the initial values $s_{\comp}$ and $r_{\comp}$ provided to them.
We also show the mean ($M$) and standard deviation ($\sigma$) values over all recorded data. The means here have not been reweighted to account for the variation in the number of sessions participants completed. However, discounting judgements from additional sessions reveals very similar results (see \textbf{S1}, Sec.~2). \cref{fig:allJudgements}c,d,g,h show individual trends for each participant. A very clear trend of underestimating the distance to the bisection point quickly emerges (\textbf{H1} validated). For \emph{size matching}, however, the trend is ambiguous (\textbf{H2} partially invalidated) but with a slight tendency towards overconstancy. The overall trends were similar for the two \emph{eye heights} for both the tasks (\textbf{H3} invalidated).

\paragraph{Data Cleaning and Outlier Removal.} 
Outliers were detected for each of the four \emph{protocol}, \emph{eye height} pairs by comparing the best-fit exponent $\alpha^*$ for each participant with those of other participants. After performing the fitting procedure (see \S~\ref{sec:testSdih}), we computed the mean and standard deviation of the natural logarithm of the exponents across all participants. Participants deviating from the mean by a $3\sigma$ threshold or higher were considered to be outliers.

Participants whose \emph{distance bisection} exponent was deemed to be an outlier for either of the two \emph{eye heights} were not considered in any of the \emph{distance bisection} analyses. Outliers for \emph{size matching} were analogously removed. This procedure led to the removal of 2 participants each for \emph{distance bisection} and \emph{size matching} analyses. Consequently, for analyses involving both the \emph{protocols}, a total of 4 participants were not considered.
In addition, one participant reported hardware issues and their data was completely discounted.
Note that \cref{fig:allJudgements} plots have been drawn after this cleaning.

From the remaining data, a very small number of trials were excluded because the participants did not press the confirm button within the time limit: a total of 12 trials (5 for \emph{size matching} and 7 for \emph{distance bisection}) over all the conditions. Thus, 3205 \emph{size matching} trials and 3203 \emph{distance bisection} trials were used for analysis.

\paragraph{Size Matching.}
The main dependent variable is the size ratio $S_{\comp} / s_{\refer}$ for each trial. Across all distances and eye-heights, the size ratio ($M \pm \sigma$) was $1.094\pm0.430$. To understand the impact of the two main factors of \emph{eye height} and \emph{reference distance}, we ran a repeated measures (RM-) ANOVA on each participant's mean response for every given pair of factor values. \emph{Reference distance} was found to be a significant factor ($F_{4,224}=4.99, p < .001$), while \emph{eye height} had a noticeable but insignificant effect ($F_{1,56}=3.76, p = .057$). No interaction effects were found. While \emph{reference distance} was a significant factor, post-hoc analysis using Tukey's Honest Significant Difference (HSD) did not reveal any significant ($p<.05$) pairwise differences. In order to assess if participants' answers were more self-consistent for certain factor levels than others, we also performed an RM-ANOVA on the standard deviation of the size ratio, which found only \emph{reference distance} to be a significant factor ($F_{4,224}=17.61, p<.001$). Tukey's HSD indicated that the distance pairs (1m,3m), (1m,5m), (1m,7m), (1m,9m), (3m,7m), (3m,9m), and (5m,9m) were significantly different.

Another effect we analyzed was ``inertia''---the tendency of initial size $s_{\comp}$ to impact the participant response $S_{\comp}$. A 1-way ANOVA found that it had a significant effect indeed: in the \emph{adult eye height}, the final size ratios were significantly larger when the initial \emph{comparison} size was larger than the \emph{reference}; $s_{\comp} > s_{\refer}$: $1.132\pm0.439$, $s_{\comp} < s_{\refer}$: $1.021 \pm 0.361$ ($F_{1,1601}= 30.6, p<.001$). The same effect was noticed for \emph{child eye height} trials: $1.146 \pm 0.481$ vs. $1.073 \pm 0.410$ ($F_{1,1600} = 10.6, p=.001$).

\paragraph{Distance Bisection.}
The main independent factor for \emph{distance bisection} was the distance ratio $R_{\comp} / r_{\refer}$ whose value over all trials was $0.469\pm0.157$. An RM-ANOVA revealed that no factor significantly affected the mean distance ratio ($p>.05$). However, a significant effect of \emph{reference distance} on the standard deviation was observed ($F_{4,224}=3.72, p=.006$). Post-hoc comparisons using Tukey's HSD did not show any significant differences.
As compared to the \emph{size matching} trials, the impact of ``inertia'' here was small. Specifically, the initial comparison distance $d_{\comp}$ did not have a significant effect on participants' responses in the \emph{adult eye height} condition, it had a small but significant effect in the \emph{child eye height} condition---distance ratio was $0.468 \pm 0.142$ when $d_{\comp} > d_{\refer}/2$ and $0.453 \pm 0.163$ when not ($F_{1,1598} = 3.91, p=.048$).


\paragraph{Learning Effects.} To confirm that our minimal aggregated feedback strategy prevented bias, we tested if participants with multiple sessions improved over time. Using paired $t$-tests for participants with two or three sessions, we observed no significant difference ($p>.10$) between the mean size ratios (\emph{size matching}) or mean distance ratios (\emph{distance bisection}) across sessions. That is, \textbf{no learning effects} were observed, as expected.

\subsection{Fitting Curves of Perceived Distance}
\label{sec:testSdih}


The size-distance invariance hypothesis (SDIH) has been used in prior work to derive perceived depth judgements from judgements of perceived size~\cite{hornsey2020size, howard2012perceiving}. In our \emph{size matching} task, participants use perceptual size matching to indicate the perceived size of the \emph{reference} by setting the \emph{comparison} size $S_{\comp}$.
The SDIH states that the ratio of perceived size $\pS$ to perceived egocentric distance $\pR$ is fixed: $\pS/\pR = \mathrm{c^p}$.

Given the actual physical size $S$ and egocentric distance $R$ for an object then, we can use the fact that their ratio is also fixed ($S/R = \mathrm{c^a}$) to invert a given judgement of perceived size, to get the perceived distance as $\pR = cR\frac{\pS}{S}$, where $\mathrm{c} = \mathrm{c^a}/\mathrm{c^p}$ is a constant modelling the ratio of an object's angular size to its perceived angular size. We assume that angular size is perceived linearly, and thus $c$ is an stimulus-independent constant.


In order to compare the perceived distance implied by the \emph{distance bisection} and the \emph{size matching} judgements, we can fit curves with the same representational degrees of freedom to trials from both and test if the parameters are similar. Similar to prior work~\cite{wagner2006geometries}, we model the perceived distance $\pR$ as an exponential function of actual distance $R$, along with a linear scaling term. That is,

\begin{equation}
    \pR = kR^\alpha \text{ with } k, \alpha \ge 0\text{.}
    \label{eq:perceivedDistance}
\end{equation}


In \emph{size matching} judgements, a participant indicates that the perceived \emph{reference} size $\pS_{\refer}$ and comparison size $\pS_{\comp}$ are equal, implying that

\[\mathrm{c^p}_{\refer} \pR_{\refer} = \mathrm{c^p}_{\comp} \pR_{\comp}.\]

Multiplying both sides by $\mathrm{c}$, we get $\mathrm{c^a}_{\refer} \pR_{\refer} = \mathrm{c^a}_{\comp} \pR_{\comp}$, which further expands to 

\[\pR_{\refer} \frac{s_{\refer}}{r_{\refer}} = \pR_{\comp}\frac{S_{\comp}}{R_{\comp}}.\]

Finally, substituting the perceived distances from Eq.~\ref{eq:perceivedDistance} and rearranging, we get the ratio of perceived distance of the \emph{comparison} object to that of the \emph{reference},

\begin{equation}
    \frac{k(R_{\comp})^\alpha}{k(r_{\refer})^\alpha} = \frac{s_{\refer}}{r_{\refer}}\frac{R_{\comp}}{S_{\comp}}\text{.}
    \label{eq:relativeDistanceSM}
\end{equation}

For accurate judgements implying an invariant visual space with no foreshortening or anti-foreshortening, the above equation should hold. Therefore, the best-fit value for $\alpha$ for \emph{size matching}, $\alpha_{SM}^*$, is obtained by minimizing the deviation from Eq.~\ref{eq:relativeDistanceSM}.

\begin{equation}
    \alpha^*_{SM} =
    \argmin_{\alpha\in(0,\infty)} \sum_i
        \left\|
            \frac{s^i_{\refer}}{r^i_{\refer}}\frac{R^i_{\comp}}{S^i_{\comp}} - \frac{(R^i_{\comp})^\alpha}{(r^i_{\refer})^\alpha}
        \right\|^2,
    \label{eq:fitSize}
\end{equation}

where the summation is over a participant's \emph{size matching} trials. Fitting a similar curve for the \emph{distance bisection} trials is relatively straightforward.

\begin{equation}
    \alpha^*_{DB} = 
    \argmin_{\alpha\in(0,\infty)} \sum_i 
        \left\|
            1 - \frac{\pR^i}{k(R^i)^\alpha}
        \right\|^2 =
    \sum_i
        \left\|
            1 - \frac{2(R^i_{\comp})^\alpha}{(r^i_{\refer})^\alpha}
        \right\|^2.
    \label{eq:fitBisect}
\end{equation}

Intuitively, note that if the curve models a participant's responses with full accuracy, then $\pR^i = k(R^i)^\alpha$ for all $i$, and the residual is zero. The minimization in \cref{eq:fitSize} penalizes relative deviation of the participant's response from the perceived distance modelled by the curve. \cref{eq:fitBisect} is similar, except that we want the modelled perceived distance to the \emph{reference} to be twice that of the \emph{comparison}. Both the minimizations are non-linear least squares problems and are efficiently solved using a trust-region based approach~\cite{coleman1996interior} available in MATLAB. We initialize with the trivial value $\alpha^*=1$ for both.


\cref{fig:teaser}c shows the curve-fitting summary results for all conditions.
The $\alpha^*$ found via \emph{size matching} judgements ($\alpha^*_{SM}$) for \emph{adult eye height} was $0.991\pm 0.185$ ($M\pm\sigma$), and for \emph{child eye height}, it was $1.012\pm 0.188$. For \emph{distance bisection} ($\alpha^*_{DB}$), it was $0.862\pm 0.219$ for \emph{adult} and $0.863\pm 0.238$ for \emph{child eye heights}. Note that exponents are fit for each participant before taking the mean and standard deviation over the participants, thus weighing single-session and multiple-session participants equally. As already suggested by analysis of the raw data, \emph{distance bisection} judgements showed a clear trend towards compression of perceptual distance space, while \emph{size matching} data indicated a mix of compression and expansion. An RM-ANOVA with dependent variable $\alpha^*$ and independent variables \emph{protocol} and \emph{eye height} found that while \emph{protocol} had a significant effect ($F_{1, 54}=15.22, p<.001$), \emph{eye height} did not ($F_{1,54}=0.007, p=.965$).


\subsection{Random Factors Effects}
\label{sec:randomEffects}


As noted in \S~\ref{sec:participants}, we collected data on a number of random factors: gender, age range, height\footnote{Height was solicited via email after the study: 45 participants responded.}, device used, 3D games experience, and 3D modelling experience. We also extracted resolution and FoV factors from device specifications. For device-independent online studies to be successful and useful for perceptual experiments, the effect of device specifications should be minimal. Therefore, we analyze the effect of each factor in isolation for the maximum statistical power, allowing us to detect and report even small deviations from the null hypotheses. For nominals (device, gender, age range), this is a 1-way ANOVA with the mean values of $\alpha^*_{DB}$ and $\alpha^*_{SM}$ for each participant as the dependents and with the random factor in consideration the only independent factor.
First considering device usage, while 11 different devices were used, we only consider the ones used by $\ge5$ participants: Valve Index (17), Oculus Quest (15), and HTC Vive (10). 1-way ANOVAs showed that neither device used nor age range were significant factors for $\alpha^*_{DB}$, but age had a significant effect on $\alpha^*_{SM}$ ($F_{3,53} = 3.46, p = .023$). Tukey's HSD revealed a pairwise difference between participants aged below 25 and those between 36 to 45 years old. Impact of gender was not analyzed due to our heavily-skewed sample.

Similarly, for numeric variables (height, resolution, and FoV) and Likert scale ordinals treated as numeric for analysis (3D games and modelling experience), we computed the correlation coefficient between each variable and the exponents $\alpha^*_{SM}$ and $\alpha^*_{DB}$, but no significant correlations were found. This is an encouraging result, indicating that an \emph{in-the-wild} strategy is viable for spatial perception experiments such as ours.


\subsection{Quantitative Analysis of Participant Strategies}
\label{sec:tracking}

The use of 6-DoF HMDs allowed us to track participants' head position and orientation. This data can help us understand participants' strategies for judging size/distance and answer questions such as ``Do they move and rotate their head a lot to use motion parallax cues?'', ``Do they look towards the ground, potentially utilizing shadows as cues?'', and ``Do they respond quickly, or not?''. To understand whether participants changed their strategies to account with changes in experiment factors, we perform RM-ANOVA with head-tracking data as dependent and experiment factors as independent variables. For simplicity, interaction effects were not included.

\ifnum\mode=1
\begin{wrapfigure}[15]{r}{\halfwidth}
	\centering
    \includegraphics[width=\linewidth]{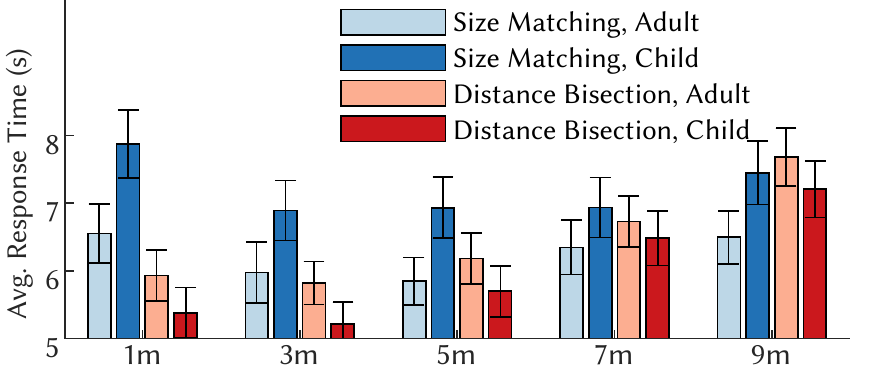}
    \caption{Average response time (with std. error) across the experiment factors. Note that the \textbf{y-axis does not start at zero}. Notice the large differences across the \emph{protocols} and \emph{eye heights} at 1m and 3m \emph{reference} distances, which shrink when the \emph{reference} is farther.}
    \label{fig:timeTaken}
\end{wrapfigure}
\fi
\paragraph{Data Collection.}
Participants' head position and orientation were recorded every 200ms. Over each trial, the difference between consecutive position samples (in $\mathbb{R}^3$) and consecutive orientation samples (in terms of angle between them) were added up to estimate the total head translation and rotation. For each participant and factor values, these two datapoints were then averaged over all associated trials to get the \emph{head translation} and \emph{head rotation} variables. In order to understand the role of shadows, we also compute \emph{looking down percentage}: the percentage of orientation samples where the cyclopean line of sight was at least 5$^\circ$\ below horizontal. We also show how the \emph{time taken} by the participant to record a response was impacted by experiment factors.



\paragraph{Change in Strategy with Experiment Factors.} RM-ANOVA indicated that the \emph{head translation} was significantly affected by the \emph{eye height} ($F_{1,53}=4.56, p=.037$)\footnote{One participant's head-tracking data was lost to a data conversion bug.}, with participants moving their heads significantly more in the \emph{adult eye height} condition ($15.5\pm22.5$ cm) as compared to \emph{child} ($13.8\pm20.9$ cm). 

\ifnum\mode=0
\begin{wrapfigure}[15]{r}{\halfwidth}
	\centering
    \includegraphics[width=\linewidth]{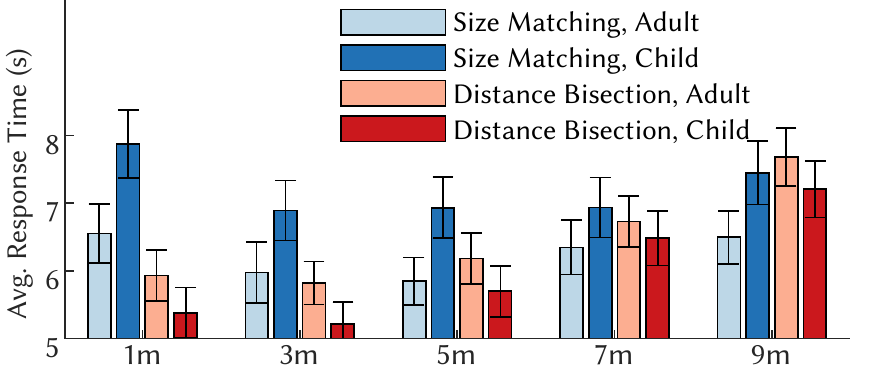}
    \caption{Average response time (with std. error) across the experiment factors. Note that the \textbf{y-axis does not start at zero}. Notice the large differences across the \emph{protocols} and \emph{eye heights} at 1m and 3m \emph{reference} distances, which shrink when the \emph{reference} is farther.}
    \label{fig:timeTaken}
\end{wrapfigure}
\fi
\emph{Head rotation} was similarly impacted by \emph{eye height} ($F_{1,53}=6.33, p=.015$), with \emph{adult} ($19.8^\circ\pm24.6^\circ$) showing significantly more rotation as compared to \emph{child} ($16.6^\circ\pm20.8^\circ$). Interestingly, \emph{looking down percentage} was significantly affected by the \emph{protocol} ($F_{1,53}=23.28, p<.001$) as well as \emph{eye height} ($F_{1,53}=4.82, p=.033$). Post-hoc comparisons showed that participants looked down much more in the \emph{distance bisection protocol} ($29.7\pm35.9\%$) as compared to \emph{size matching} ($9.0\pm18.1\%$), and in the \emph{adult eye height} condition ($21.8\pm31.8\%$) as compared to \emph{child} ($17.0\pm28.4\%$). Note that the effect sizes are larger for the difference between \emph{protocols}.
The \emph{time taken} measure saw a significant effect of \emph{reference distance} ($F_{4,216}=20.9, p<.001$). Participants took the most time of 7.21 s for the 9m distance, as expected, but the least time was taken for the 3m distance: 5.97 s. \emph{Time taken} for the 9m distance significantly differed from other distance values, and a significant difference was also observed between (1m, 3m), (3m, 5m), and (5m, 7m) pairs. \cref{fig:timeTaken} summarizes the results for the \emph{time taken} variable.


\paragraph{Impact of Strategy on Perceptual Judgements.} In order to understand the impact of participants' strategy (encoded in the variables described above) on their perceptual judgements, we compute correlations of these variables with the distance perception exponent $\alpha^*$ for all four combinations of \emph{protocol} and \emph{eye height}. Multiple significant but weak correlations were found for $\alpha^*_{DB}$ for \emph{adult eye height}---with \emph{head translation} ($r(54)=.29, p=.030$), \emph{head rotation} ($r(54)=.29, p=.030$), and \emph{looking down percentage} ($r(54)=.27, p=.044$).

Lastly, we had asked participants for introspective reports of how they judged size/distance in each condition.

\paragraph{Participants' Introspective Reports}
\label{sec:introspect}

We sought introspective reports from participants in two forms: freeform and structured. In order to not bias participants with our own presumptions, we first asked them for freeform feedback on the cues they used in both the \emph{protocols}, and if the \emph{eye height} impacted the cues they used. In a later section of the survey form (see supplemental \textbf{S3}), participants answered structured (multiple-choice) questions on the cues utilized in the two \emph{protocols}.

Participants were surprisingly accurate in certain aspects. For example, those who mentioned the use of motion parallax in the structured questions exhibited much larger \emph{head translation}: mean value was 21.6 cm vs. 11.0 cm for \emph{size matching} and 17.5 cm vs. 10.7 cm for \emph{distance bisection}. An ANOVA found the former distinction to be insignificant ($F_{1,55}=3.00, p=.089$), but the latter significant ($F_{1,55}=6.26, p=.015$). \emph{Head rotation} showed a similar trend---\emph{size matching}: 21.7$^\circ$\ vs. 14.1$^\circ$, \emph{distance bisection}: 24.2$^\circ$\ vs. 14.7$^\circ$. The former was again found to be statistically insignificant ($F_{1,55}=2.79, p=.100$) but the latter was significant ($F_{1,55}=5.69, p=.021$). Similarly, participants who reported using shadows as a cue were more likely to spend time \emph{looking down}: 13.3\% vs. 6.6\% ($F_{1,55}=2.66,p=.108$) for \emph{size matching} and 35.9\% vs. 21.5\% ($F_{1,55}=2.30,p=.135$) for \emph{distance bisection}.

The structured reports helped identify a few other interesting cue usage patterns. For instance, many participants tried to measure absolute egocentric size/distance in standard units (\emph{SM}: 25, \emph{DB}: 25), and a smaller but still significant number used relative size/distance cues such as their own body parts (\emph{SM}: 16, \emph{DB}: 6). A number of participants (\emph{SM}: 12, \emph{DB}: 16) felt that the speed of the continuous change in size/distance of the \emph{comparison} was a useful cue. This is in spite of our efforts to minimize the usefulness of this cue: we used an acceleration term so that the speed was a function of the amount of time the participant held the thumbstick (used to increase or decrease the size/distance) in place.

Finally, subjective analysis of participants' freeform reports revealed some interesting insights. 34 out of 55 participants (whose data was used for both \emph{protocols}) reported that the \emph{eye height} did not influence their judgements at all, while another 4 felt that it had minimal impact. Most (11) of the participants who felt that it affected their judgements noted that the shadows were harder to see or otherwise different in one of the \emph{eye heights}. Surprisingly, very few (4) noted any additional difficulty in the \emph{child height} condition they had little experience with in real life. One participant commented that they were \pquote{very used to not looking at the floor for the sake of avoiding motion sickness}, so the \emph{eye height} did not influence them.

Some participants relied on prior experience with VR games to estimate physical sizes. One said, \pquote{I would think back to all the times I see [sic] an object in the distance and come close from other games I have played.}, while another \pquote{channeled a couple of years of VR Beat Saber and Synth Riders as those involve basically constant estimation of distance.} Other participants mentioned that the ability to continuously move the \emph{comparison} back and forth was helpful for \emph{distance bisection}: \pquote{I moved the controlled cube back and forth to try and understand how far away the reference cube was\ldots}. One participant talked about using the two monocular views separately to get a better estimate of depth: \pquote{I just tried to compare the sizes while alternating which eye was open.}

%% file: 6_discussion.tex
We discuss the implications of our work on visual perception in VR, and on remote perceptual experiments.

\subsection{Implications on VR Perception Understanding}
\label{sec:perceptDiscussion}


Our size perception experiment used a perceptual matching protocol inspired by Holway and Boring's classic experiment~\shortcite{holway1941determinants}. While our results were similar to the overall trend of slight size overconstancy they noted, we observed more variation and ambiguity---with individuals exhibiting both underconstancy and overconstancy.
Other classic experiments~\cite{epstein1969size} and more recent VR experiments~\cite{peillard2019virtual} have studied restricted viewing conditions. Aiming for ecological validity, we employed scenarios closer to real-world usage: binocular viewing, relatively unrestricted head movement, lighting, and a textured ground plane. In this aspect, the experiments closest to ours might be Kenyon et al.~\shortcite{kenyon2007size}, Hornsey et al.~\shortcite{hornsey2020size}, and Kelly et al.~\cite{kelly2017perceived}. 

Kenyon et al.~\shortcite{kenyon2007size} looked at size constancy in a CAVE environment, asking participants to resize a far-away familiar object while the real object was also placed near them, and observed that participants made the distant comparison larger than the reference across the range of tested distances (0.6--2.4m). The other two studies~\cite{hornsey2020size,kelly2017perceived} used modern HMDs and a familiar-sized object that the participants could touch by hand for reference. They also observed a general trend of size underconstancy. Note that our protocol is reversed: the reference is far away while the participant-controlled comparison is nearby. Therefore, while the previous studies observed size-underconstancy in VR, we found a slight overconstancy trend. One possible explanation to this contrary effect may be the shape of the stimulus. Wagner's survey on real-world size constancy~\cite{wagner2012sensory} points out that size perception may be related to object orientation: frontally-oriented (perpendicular to the floor and facing the participant) stimuli were commonly associated with overconstancy and flat (parallel to the floor) stimuli with underconstancy. In comparison to the spherical~\cite{kelly2017perceived, hornsey2020size} and cylindrical~\cite{kenyon2007size} objects used in prior VR studies, our cube-shaped stimuli have a more noticeable frontal face.

Wagner~\shortcite{wagner2012sensory} notes that an important factor in size perception experiments is the instruction provided to the subjects. While we aimed for \emph{objective instructions}---asking participants to match the physical (3D) size, some participants may still have been influenced by the apparent, or retinal, size, leading to size underconstancy. On the contrary, some may have overestimated the perspective foreshortening effect, leading to overconstancy. 

Our bisection protocol is similar to bisection studies in the real world~\cite{lappin2006environmnet,bodenheimer2007distance} as well as those in virtual environments~\cite{bodenheimer2007distance}. While Lappin et al.~\shortcite{lappin2006environmnet} noted an anomalous report of overestimation of the distance to the bisection point, most studies (especially in VR) report underestimation, similar to ours. Moreover, egocentric distance judgements made using very different protocols have typically suggested distance underconstancy as well~\cite[survey]{renner2013perception}. The latter lends further cadence to the values of $\alpha^*_{DB}$ our curve-fitting suggested: in either \emph{eye height}, bisection judgements suggested a shrinkage of perceived distance towards the observer.

Similar to our exponential curve fitting, Murgia and Sharkey~\cite{murgia2009estimation} reported mean exponent values of $0.830$ in an environment with poor cues, and $0.975$ with one with rich cues, albeit in a CAVE system. Our environment rendering falls somewhere between those two: we render a floor, but no walls. Thus, the exponent $0.896$ we observed in our bisection protocol is more similar to their results, than the value $1.083$ suggested in the size matching protocol. This dichotomy is especially interesting since their results were obtained via a perceptual size matching protocol.


While our experiments do not intend to test the validity of SDIH, we observed a weak correlation between $\alpha^*_{DB}$ and $\alpha^*_{SM}$, which is worthwhile to put in context of Ross's investigation of SDIH alternatives~\cite{ross2003levels}. Of particular interest is his suggestion of a ``perceptual SDIH'', which breaks the classical SDIH assumption of accurately perceived angular size. In a future experiment, Wagner's \emph{angular size instruction} can be applied to gauge the accuracy of angular size perception, by using a perceptual matching protocol similar to Kaneko and Uchikawa~\shortcite{kaneko1997perceived}.

Contrary to our initial hypothesis, comparisons across the two eye heights did not reveal any noticable differences. This is in contrast to Corujeira and Ian's study~\shortcite{corujeira2013stereocopic} where observers were \emph{better} at judging egocentric distances in the lower eye-height (20 cm from the floor) condition. However, this comparison should be taken with a grain of salt: they position objects on the ground and define distance as the walking distance to the percept. Thus, lower the eye-height, better the correspondence between the 3D egocentric distance and the walking distance. Leyrer et al.~\shortcite{leyrer2011influence}, who studied virtually altered eye heights in VR similar to us, also found that the foreshortening effect was smaller when eye height was closer to the ground plane, but they also position stimuli on the ground and define distance as the 2D distance projected to the floor.

Finally, our quantitative observation of the minimal impact of head motion on size/distance judgements in VR confirms Hornsey et al.'s informal observation~\shortcite{hornsey2020size}. We further discuss Hornsey et al. and other relatively-uncontrolled perceptual experiments below.

\subsection{Conducting Remote Perceptual Studies}
\label{sec:remoteDiscussion}




Our remote unsupervised methodology allowed us to quickly recruit 60 participants within 5 days. Judging by comments and feedback on our advertisements, the most important source of participants was Reddit. VR-focussed communities on Reddit showed great enthusiasm for the study, reconfirming previous work hailing the platform as a useful community for recruiting participants~\cite{shatz2017fast}. To engage participants, we used a combination of gamification and intrinsic as well as extrinsic motivation. Specifically, participants were told that the study allowed them to test their size and depth perception abilities in VR and contribute to science (\emph{intrinsic motivation}), while at the same time receiving a monetary compensation (\emph{extrinsic motivation}), and getting to see their score on a leaderboard with a chance to get a higher compensation (\emph{gamification} and \emph{extrinsic motivation}).

We believe that the gamification elements ensured increased user engagement, resulting in high-quality data in our experiment. Furthermore, we did not see any evidence for malicious behaviour. This is in contrast to microtask-based crowdsourced studies on platforms such as Mechanical Turk, where testing for malicious users is commonplace~\cite{snow2008cheap}. Prior work has also shown that dropping monetary compensation altogether still allows successful unsupervised studies with a large number of participants~\cite{huber2020conducting}. Outside of perception research, many well-known scientific projects have also used gamification and intrinsic motivation to engage the target audience~\cite{curtis2015motivation,huynh2016analyzing}. This combination is an exciting exploration avenue for future VR perception research. 

Our study included participants from 19 different countries and a large variety of HMDs. While this diversity is encouraging, access to 6-DoF VR devices continues to be dominated by men in rich countries. This demographics bias may have affected some of our results. We hope that this bias in VR ownership will go down with lowering hardware costs democratizing access in low- and medium-income countries, and that research on the unique challenges faced by female VR users~\cite{stanney2020virtual} reduces the gender bias in the future. Some bias towards participants in the US and Canada may also have occurred due to better availability of many VR devices in the region and the fact that we posted our recruitment calls in English-language forums. However, even with the currently skewed VR ownership demographics, remote VR studies are still an invaluable tool for perception research and offer researchers access to a more diverse audience than their immediate local community.

Finally, the variety of headsets used by our participants allowed us to test the impact of the device on perceptual judgements. Our results echo the partially-uncontrolled experiment of Hornsey et al.~\cite{hornsey2020size}, who tested size and shape judgements with two headsets and did not observe any significant differences between the two. However, it is unclear if this generalizes to perceptual judgements unexplored by our or their work. Previous work has noted a significant effect of hardware factors such as FoV~\cite{jones2012comparability} and IPD~\cite{bruder2012analyzing}. Further exploration is needed to understand which perceptual studies may not be amenable to uncontrolled methodologies such as ours.

%% file: 7_conclusion.tex
We conducted a remote, fully unsupervised, perceptual study to understand size and distance perception in Virtual Reality. To recruit study participants and to keep them engaged through the monotonous tasks, we combined an extrinsic monetary reward with gamification and an appeal to their intrinsic motivation. Ours is the first fully-uncontrolled VR distance/size perception study, aimed at ecological validity by collecting data from a diverse set of participants and VR hardware. Unfortunately, current VR ownership trends reduced participant diversity in some aspects, most notably gender.

We compare the results obtained via a distance bisection protocol with distance perception implied by a size matching protocol via the size-distance invariance hypothesis (SDIH). While the former indicated a clear trend of distance underestimation, the latter showed a very weak overestimation trend. 
We also investigated the influence of a rendered floor, simulating different eye heights, but noticed no significant effect. We distilled participants' judgements into exponential curves of perceived egocentric distance expressed as a function of actual distance. Such curves can be used to manipulate virtual worlds to improve task performance in performance-critical applications such as VR flight simulators and industrial training as well as in 3D judgement--heavy games (\cref{fig:teaser}d).

To aid future research, we will \textbf{release the source code for our experiment and analysis, as well as anonymized data}. The study executable is included (see \textbf{S3}) for reviewers to test and analyze.

\paragraph{Potential Applications and Future Work.}
\label{sec:futureWork}
In the future, we would like to systematically explore the gamification aspects of our work, modifying leaderboard size and monetary rewards parameters, and exploring a deeper coupling of gamification with the perceptual judgement tasks. Directly comparing intrinsic motivations with microtasks is also an interesting avenue. On the other hand, we are also inspired by recent work by Huber and Gajos~\shortcite{huber2020conducting} that advocates for conducting large scale studies without any monetary rewards.

Finally, we are very interested in applying the perceptual distance curves to real-world applications. Perceptual judgement tasks, such as distance bisection, could be used as a ``pre-calibration'' step for modifying the rendered world in performance-critical applications, similar to gamma calibration on monitors. In games, such a calibration can be utilized for improving performance as well, or a sneaky way to make the game more difficult for users by applying the inverse transformation to the rendered distance.

%% file: arxiv_vrperception_main.bbl
\begin{thebibliography}{10}

\bibitem{beall1995absolute}
A.~C. Beall, J.~M. Loomis, J.~W. Philbeck, and T.~G. Fikes.
\newblock {Absolute motion parallax weakly determines visual scale in real and
  virtual environments}.
\newblock In B.~E. Rogowitz and J.~P. Allebach, eds., {\em Human Vision, Visual
  Processing, and Digital Display VI}, vol. 2411, pp. 288 -- 297. International
  Society for Optics and Photonics, SPIE, 1995. doi: {{%
10\hspace{.1pt}\discretionary{.}{%
}{.}\hspace{.4pt}1117\discretionary{/}{%
}{/}12\hspace{.1pt}\discretionary{.}{%
}{.}\hspace{.4pt}207547}}


\bibitem{bingham2001accommodation}
G.~P. Bingham, A.~Bradley, M.~Bailey, and R.~Vinner.
\newblock Accommodation, occlusion, and disparity matching are used to guide
  reaching: A comparison of actual versus virtual environments.
\newblock {\em Journal of experimental psychology: human perception and
  performance}, 27(6):1314, 2001.

\bibitem{bodenheimer2007distance}
B.~Bodenheimer, J.~Meng, H.~Wu, G.~Narasimham, B.~Rump, T.~P. McNamara, T.~H.
  Carr, and J.~J. Rieser.
\newblock Distance estimation in virtual and real environments using bisection.
\newblock In {\em Proceedings of the 4th Symposium on Applied Perception in
  Graphics and Visualization}, pp. 35--40, 2007.

\bibitem{bruder2012analyzing}
G.~Bruder, A.~Pusch, and F.~Steinicke.
\newblock Analyzing effects of geometric rendering parameters on size and
  distance estimation in on-axis stereographics.
\newblock In {\em Proceedings of the ACM Symposium on Applied Perception}, pp.
  111--118, 2012.

\bibitem{bruder2011tuning}
G.~Bruder, F.~Steinicke, P.~Wieland, and M.~Lappe.
\newblock Tuning self-motion perception in virtual reality with visual
  illusions.
\newblock {\em IEEE Transactions on Visualization and Computer Graphics},
  18(7):1068--1078, 2011.

\bibitem{coleman1996interior}
T.~F. Coleman and Y.~Li.
\newblock An interior trust region approach for nonlinear minimization subject
  to bounds.
\newblock {\em SIAM Journal on optimization}, 6(2):418--445, 1996.

\bibitem{corujeira2013stereocopic}
J.~G.~P. Corujeira and I.~Oakley.
\newblock Stereoscopic egocentric distance perception: The impact of eye height
  and display devices.
\newblock In {\em Proceedings of the ACM Symposium on Applied Perception}, SAP
  '13, p. 23–30. Association for Computing Machinery, New York, NY, USA,
  2013. doi: {{%
10\hspace{.1pt}\discretionary{.}{%
}{.}\hspace{.4pt}1145\discretionary{/}{%
}{/}2492494\hspace{.1pt}\discretionary{.}{%
}{.}\hspace{.4pt}2492509}}


\bibitem{curtis2015motivation}
V.~Curtis.
\newblock Motivation to participate in an online citizen science game: A study
  of foldit.
\newblock {\em Science Communication}, 37(6):723--746, 2015.

\bibitem{cutting2003reconceiving}
J.~Cutting.
\newblock {\em Reconceiving perceptual space}, pp. 215--238.
\newblock MIT Press, Cambridge, MA, USA, 01 2003.

\bibitem{day1989exorcize}
R.~Day and T.~E. Parks.
\newblock To exorcize a ghost from the perceptual machine.
\newblock {\em The moon illusion}, pp. 343--350, 1989.

\bibitem{eby1995perceptual}
D.~W. Eby and M.~L. Braunstein.
\newblock The perceptual flattening of three-dimensional scenes enclosed by a
  frame.
\newblock {\em Perception}, 24(9):981--993, 1995.

\bibitem{eggleston1996virtual}
R.~G. Eggleston, W.~P. Janson, and K.~A. Aldrich.
\newblock Virtual reality system effects on size-distance judgements in a
  virtual environment.
\newblock In {\em Proceedings of the IEEE 1996 Virtual Reality Annual
  International Symposium}, pp. 139--146. IEEE, 1996.

\bibitem{epstein1963attitudes}
W.~Epstein.
\newblock Attitudes of judgment and the size-distance invariance hypothesis.
\newblock {\em Journal of experimental Psychology}, 66(1):78, 1963.

\bibitem{epstein1969size}
W.~Epstein and A.~A. Landauer.
\newblock Size and distance judgments under reduced conditions of viewing.
\newblock {\em Perception \& Psychophysics}, 6(5):269--272, 1969.

\bibitem{epstein1961current}
W.~Epstein, J.~Park, and A.~Casey.
\newblock The current status of the size-distance hypotheses.
\newblock {\em Psychological Bulletin}, 58(6):491, 1961.

\bibitem{foley2004visual}
J.~M. Foley, N.~P. Ribeiro-Filho, and J.~A. Da~Silva.
\newblock Visual perception of extent and the geometry of visual space.
\newblock {\em Vision Research}, 44(2):147--156, 2004.

\bibitem{galais2014you}
C.~Galais and E.~Anduiza.
\newblock ``{You} cheated on me!'' {Causes} and consequences of cheating in
  online surveys.
\newblock In {\em Visions in Methodology (VIM)}. McMaster University, Hamilton,
  Canada, 2014.

\bibitem{gehlbach2015many}
H.~Gehlbach, G.~Marietta, A.~M. King, C.~Karutz, J.~N. Bailenson, and C.~Dede.
\newblock Many ways to walk a mile in another’s moccasins: Type of social
  perspective taking and its effect on negotiation outcomes.
\newblock {\em Computers in Human Behavior}, 52:523--532, 2015.

\bibitem{gilinsky1951perceived}
A.~S. Gilinsky.
\newblock Perceived size and distance in visual space.
\newblock {\em Psychological review}, 58(6):460, 1951.

\bibitem{glennerster2006humans}
A.~Glennerster, L.~Tcheang, S.~J. Gilson, A.~W. Fitzgibbon, and A.~J. Parker.
\newblock Humans ignore motion and stereo cues in favor of a fictional stable
  world.
\newblock {\em Current Biology}, 16(4):428 -- 432, 2006. doi: {{%
10\hspace{.1pt}\discretionary{.}{%
}{.}\hspace{.4pt}1016\discretionary{/}{%
}{/}j\hspace{.1pt}\discretionary{.}{%
}{.}\hspace{.4pt}cub\hspace{.1pt}\discretionary{.}{%
}{.}\hspace{.4pt}2006\hspace{.1pt}\discretionary{.}{%
}{.}\hspace{.4pt}01\hspace{.1pt}\discretionary{.}{%
}{.}\hspace{.4pt}019}}


\bibitem{gogel1987familiar}
W.~Gogel and J.~A. Da~Silva.
\newblock Familiar size and the theory of off-sized perceptions.
\newblock {\em Perception \& psychophysics}, 41:318--28, 05 1987. doi: {{%
10\hspace{.1pt}\discretionary{.}{%
}{.}\hspace{.4pt}3758\discretionary{/}{%
}{/}BF03208233}}


\bibitem{gould2016diminished}
S.~J. Gould, A.~L. Cox, and D.~P. Brumby.
\newblock Diminished control in crowdsourcing: an investigation of crowdworker
  multitasking behavior.
\newblock {\em ACM Transactions on Computer-Human Interaction (TOCHI)},
  23(3):1--29, 2016.

\bibitem{gruber1954relation}
H.~E. Gruber.
\newblock The relation of perceived size to perceived distance.
\newblock {\em The American Journal of Psychology}, 67(3):411--426, 1954.

\bibitem{haber2001independence}
R.~N. Haber and C.~A. Levin.
\newblock The independence of size perception and distance perception.
\newblock {\em Perception \& psychophysics}, 63(7):1140--1152, 2001.

\bibitem{hata2017glimpse}
K.~Hata, R.~Krishna, L.~Fei-Fei, and M.~S. Bernstein.
\newblock A glimpse far into the future: Understanding long-term crowd worker
  quality.
\newblock In {\em Proceedings of the 2017 ACM Conference on Computer Supported
  Cooperative Work and Social Computing}, pp. 889--901, 2017.

\bibitem{hirth2011cost}
M.~Hirth, T.~Ho{\ss}feld, and P.~Tran-Gia.
\newblock Cost-optimal validation mechanisms and cheat-detection for
  crowdsourcing platforms.
\newblock In {\em 2011 fifth international conference on innovative mobile and
  internet services in ubiquitous computing}, pp. 316--321. IEEE, 2011.

\bibitem{hoffman2008vergence}
D.~M. Hoffman, A.~R. Girshick, K.~Akeley, and M.~S. Banks.
\newblock Vergence--accommodation conflicts hinder visual performance and cause
  visual fatigue.
\newblock {\em Journal of vision}, 8(3):33--33, 2008.

\bibitem{holway1941determinants}
A.~H. Holway and E.~G. Boring.
\newblock Determinants of apparent visual size with distance variant.
\newblock {\em The American Journal of Psychology}, 54(1):21--37, 1941.

\bibitem{hornsey2020size}
R.~L. Hornsey, P.~B. Hibbard, and P.~Scarfe.
\newblock Size and shape constancy in consumer virtual reality.
\newblock {\em Behavior research methods}, 52(4):1587—1598, August 2020. doi:
  {{%
10\hspace{.1pt}\discretionary{.}{%
}{.}\hspace{.4pt}3758\discretionary{/}{%
}{/}s13428\discretionary{%
}{-}{-}019\discretionary{%
}{-}{-}01336\discretionary{%
}{-}{-}9}}


\bibitem{howard2012perceiving}
I.~Howard.
\newblock {\em Perceiving in Depth. Volume 3. Other mechanisms of depth
  perception}.
\newblock Oxford University Press, 03 2012.

\bibitem{huber2020conducting}
B.~Huber and K.~Z. Gajos.
\newblock Conducting online virtual environment experiments with uncompensated,
  unsupervised samples.
\newblock {\em Plos one}, 15(1):e0227629, 2020.

\bibitem{huynh2016analyzing}
D.~Huynh, L.~Zuo, and H.~Iida.
\newblock Analyzing gamification of ``duolingo'' with focus on its course
  structure.
\newblock In {\em International Conference on Games and Learning Alliance}, pp.
  268--277. Springer, 2016.

\bibitem{jenkin1957effects}
N.~Jenkin.
\newblock Effects of varied distance on short-range size judgments.
\newblock {\em Journal of experimental psychology}, 54(5):327, 1957.

\bibitem{jenkin1959relationship}
N.~Jenkin.
\newblock A relationship between increments of distance and estimates of
  objective size.
\newblock {\em The American Journal of Psychology}, 72(3):345--363, 1959.

\bibitem{jones2012comparability}
J.~A. Jones, E.~A. Suma, D.~M. Krum, and M.~Bolas.
\newblock Comparability of narrow and wide field-of-view head-mounted displays
  for medium-field distance judgments.
\newblock In {\em Proceedings of the ACM Symposium on Applied Perception}, pp.
  119--119, 2012.

\bibitem{kaneko1997perceived}
H.~Kaneko and K.~Uchikawa.
\newblock Perceived angular and linear size: the role of binocular disparity
  and visual surround.
\newblock {\em Perception}, 26(1):17--27, 1997.

\bibitem{kelly2017perceived}
J.~W. Kelly, L.~A. Cherep, and Z.~D. Siegel.
\newblock Perceived space in the {HTC} {Vive}.
\newblock {\em ACM Transactions on Applied Perception (TAP)}, 15(1):1--16,
  2017.

\bibitem{kenyon2007size}
R.~V. Kenyon, D.~Sandin, R.~C. Smith, R.~Pawlicki, and T.~Defanti.
\newblock Size-constancy in the cave.
\newblock {\em Presence: Teleoperators and Virtual Environments},
  16(2):172--187, 2007.

\bibitem{komarov2013crowdsourcing}
S.~Komarov, K.~Reinecke, and K.~Z. Gajos.
\newblock Crowdsourcing performance evaluations of user interfaces.
\newblock In {\em Proceedings of the SIGCHI conference on human factors in
  computing systems}, pp. 207--216, 2013.

\bibitem{koyama2014crowd}
Y.~Koyama, D.~Sakamoto, and T.~Igarashi.
\newblock Crowd-powered parameter analysis for visual design exploration.
\newblock In {\em Proceedings of the 27th Annual ACM Symposium on User
  Interface Software and Technology}, UIST '14, p. 65–74. Association for
  Computing Machinery, New York, NY, USA, 2014. doi: {{%
10\hspace{.1pt}\discretionary{.}{%
}{.}\hspace{.4pt}1145\discretionary{/}{%
}{/}2642918\hspace{.1pt}\discretionary{.}{%
}{.}\hspace{.4pt}2647386}}


\bibitem{koyama2017sequential}
Y.~Koyama, I.~Sato, D.~Sakamoto, and T.~Igarashi.
\newblock Sequential line search for efficient visual design optimization by
  crowds.
\newblock {\em ACM Trans. Graph.}, 36(4), July 2017. doi: {{%
10\hspace{.1pt}\discretionary{.}{%
}{.}\hspace{.4pt}1145\discretionary{/}{%
}{/}3072959\hspace{.1pt}\discretionary{.}{%
}{.}\hspace{.4pt}3073598}}


\bibitem{lappin2006environmnet}
J.~Lappin, A.~Shelton, and J.~Rieser.
\newblock Environmental context influences visually perceived distance.
\newblock {\em Perception \& psychophysics}, 68:571--81, 06 2006. doi: {{%
10\hspace{.1pt}\discretionary{.}{%
}{.}\hspace{.4pt}3758\discretionary{/}{%
}{/}BF03208759}}


\bibitem{leyrer2011influence}
M.~Leyrer, S.~A. Linkenauger, H.~H. B\"{u}lthoff, U.~Kloos, and B.~Mohler.
\newblock The influence of eye height and avatars on egocentric distance
  estimates in immersive virtual environments.
\newblock In {\em Proceedings of the ACM SIGGRAPH Symposium on Applied
  Perception in Graphics and Visualization}, APGV '11, p. 67–74. Association
  for Computing Machinery, New York, NY, USA, 2011. doi: {{%
10\hspace{.1pt}\discretionary{.}{%
}{.}\hspace{.4pt}1145\discretionary{/}{%
}{/}2077451\hspace{.1pt}\discretionary{.}{%
}{.}\hspace{.4pt}2077464}}


\bibitem{loomis2003visual}
J.~Loomis and J.~Knapp.
\newblock Visual perception of egocentric distance in real and virtual
  environments.
\newblock In L.~Hettinger and M.~Hass, eds., {\em Virtual and adaptive
  environments: Applications, implications, and human performance issues}, pp.
  21--46. Lawrence Erlbaum Associates Publishers, Mahwah, NJ, USA, 2003.

\bibitem{loomis1992visual}
J.~M. Loomis, J.~A. Da~Silva, N.~Fujita, and S.~S. Fukusima.
\newblock Visual space perception and visually directed action.
\newblock {\em Journal of Experimental Psychology: Human Perception and
  Performance}, 18(4):906, 1992.

\bibitem{loomis1996visual}
J.~M. Loomis, J.~A. Da~Silva, J.~W. Philbeck, and S.~S. Fukusima.
\newblock Visual perception of location and distance.
\newblock {\em Current Directions in Psychological Science}, 5(3):72--77, 1996.

\bibitem{looyestyn2017does}
J.~Looyestyn, J.~Kernot, K.~Boshoff, J.~Ryan, S.~Edney, and C.~Maher.
\newblock {Does gamification increase engagement with online programs? A
  systematic review}.
\newblock {\em PloS one}, 12(3):e0173403, 2017.

\bibitem{ma2018international}
X.~Ma, M.~Cackett, L.~Park, E.~Chien, and M.~Naaman.
\newblock {Web-Based VR Experiments Powered by the Crowd}.
\newblock In {\em Proceedings of the 2018 World Wide Web Conference}, WWW '18,
  p. 33–43. International World Wide Web Conferences Steering Committee,
  Republic and Canton of Geneva, CHE, 2018. doi: {{%
10\hspace{.1pt}\discretionary{.}{%
}{.}\hspace{.4pt}1145\discretionary{/}{%
}{/}3178876\hspace{.1pt}\discretionary{.}{%
}{.}\hspace{.4pt}3186034}}


\bibitem{mason2012conducting}
W.~Mason and S.~Suri.
\newblock Conducting behavioral research on amazon’s mechanical turk.
\newblock {\em Behavior research methods}, 44(1):1--23, 2012.

\bibitem{mottelson2017virtual}
A.~Mottelson and K.~Hornb{\ae}k.
\newblock Virtual reality studies outside the laboratory.
\newblock In {\em Proceedings of the 23rd acm symposium on virtual reality
  software and technology}, pp. 1--10, 2017.

\bibitem{murgia2009estimation}
A.~Murgia and P.~M. Sharkey.
\newblock Estimation of distances in virtual environments using size constancy.
\newblock {\em International Journal of Virtual Reality}, 8(1):67--74, 2009.

\bibitem{naceri2015depth}
A.~Naceri, A.~Moscatelli, and R.~Chellali.
\newblock Depth discrimination of constant angular size stimuli in action
  space: role of accommodation and convergence cues.
\newblock {\em Frontiers in Human Neuroscience}, 9:511, 2015. doi: {{%
10\hspace{.1pt}\discretionary{.}{%
}{.}\hspace{.4pt}3389\discretionary{/}{%
}{/}fnhum\hspace{.1pt}\discretionary{.}{%
}{.}\hspace{.4pt}2015\hspace{.1pt}\discretionary{.}{%
}{.}\hspace{.4pt}00511}}


\bibitem{norman1996visual}
J.~F. Norman, J.~T. Todd, V.~J. Perotti, and J.~S. Tittle.
\newblock The visual perception of three-dimensional length.
\newblock {\em Journal of Experimental Psychology: Human Perception and
  Performance}, 22(1):173, 1996.

\bibitem{peillard2019virtual}
E.~{Peillard}, T.~{Thebaud}, J.~{Normand}, F.~{Argelaguet}, G.~{Moreau}, and
  A.~{Lécuyer}.
\newblock Virtual objects look farther on the sides: The anisotropy of distance
  perception in virtual reality.
\newblock In {\em 2019 IEEE Conference on Virtual Reality and 3D User
  Interfaces (VR)}, pp. 227--236, 2019.

\bibitem{perlin2002improving}
K.~Perlin.
\newblock Improving noise.
\newblock In {\em Proceedings of the 29th annual conference on Computer
  graphics and interactive techniques}, pp. 681--682. ACM, New York, NY, USA,
  2002.

\bibitem{philbeck1997comparison}
J.~W. Philbeck and J.~M. Loomis.
\newblock Comparison of two indicators of perceived egocentric distance under
  full-cue and reduced-cue conditions.
\newblock {\em Journal of Experimental Psychology: Human Perception and
  Performance}, 23(1):72, 1997.

\bibitem{purdy1955distance}
J.~Purdy and E.~J. Gibson.
\newblock Distance judgment by the method of fractionation.
\newblock {\em Journal of Experimental Psychology}, 50(6):374, 1955.

\bibitem{ragozin2020mazerunvr}
K.~Ragozin, K.~Kunze, K.~Marky, and Y.~S. Pai.
\newblock {MazeRunVR: An Open Benchmark for VR Locomotion Performance,
  Preference and Sickness in the Wild}.
\newblock In {\em Extended Abstracts of the 2020 CHI Conference on Human
  Factors in Computing Systems}, pp. 1--8, 2020.

\bibitem{rakuten2019mobile}
{Rakuten Intelligence}.
\newblock The brief blog: Mobile dominates the online reality of virtual
  reality sales.
\newblock
  \url{https://www.rakutenintelligence.com/blog/2016/virtual-reality-mostly-mobile},
  2019.

\bibitem{reinecke2015labinthewild}
K.~Reinecke and K.~Z. Gajos.
\newblock {LabintheWild}: {Conducting} large-scale online experiments with
  uncompensated samples.
\newblock In {\em Proceedings of the 18th ACM conference on computer supported
  cooperative work \& social computing}, pp. 1364--1378, 2015.

\bibitem{renner2013perception}
R.~S. Renner, B.~M. Velichkovsky, and J.~R. Helmert.
\newblock The perception of egocentric distances in virtual environments-a
  review.
\newblock {\em ACM Computing Surveys (CSUR)}, 46(2):1--40, 2013.

\bibitem{rieser1990visual}
J.~J. Rieser, D.~H. Ashmead, C.~R. Talor, and G.~A. Youngquist.
\newblock Visual perception and the guidance of locomotion without vision to
  previously seen targets.
\newblock {\em Perception}, 19(5):675--689, 1990.

\bibitem{ross2003levels}
H.~E. Ross.
\newblock Levels of processing in the size-distance paradox.
\newblock In {\em Levels of perception}, pp. 149--168. Springer, 2003.

\bibitem{sasaki2019crowdsourcing}
K.~Sasaki and Y.~Yamada.
\newblock Crowdsourcing visual perception experiments: a case of contrast
  threshold.
\newblock {\em PeerJ}, 7, 2019.

\bibitem{shatz2017fast}
I.~Shatz.
\newblock Fast, free, and targeted: Reddit as a source for recruiting
  participants online.
\newblock {\em Social Science Computer Review}, 35(4):537--549, 2017. doi: {{%
10\hspace{.1pt}\discretionary{.}{%
}{.}\hspace{.4pt}1177\discretionary{/}{%
}{/}0894439316650163}}


\bibitem{smith1953methodological}
W.~M. Smith.
\newblock A methodological study of size-distance perception.
\newblock {\em The Journal of Psychology}, 35(1):143--153, 1953.

\bibitem{snow2008cheap}
R.~Snow, B.~O'Connor, D.~Jurafsky, and A.~Y. Ng.
\newblock Cheap and fast---but is it good? evaluating non-expert annotations
  for natural language tasks.
\newblock In {\em Proceedings of the Conference on Empirical Methods in Natural
  Language Processing}, EMNLP '08, p. 254–263. Association for Computational
  Linguistics, USA, 2008.

\bibitem{stanney2020virtual}
K.~Stanney, C.~Fidopiastis, and L.~Foster.
\newblock Virtual reality is sexist: But it does not have to be.
\newblock {\em Frontiers in Robotics and AI}, 7:4, 2020. doi: {{%
10\hspace{.1pt}\discretionary{.}{%
}{.}\hspace{.4pt}3389\discretionary{/}{%
}{/}frobt\hspace{.1pt}\discretionary{.}{%
}{.}\hspace{.4pt}2020\hspace{.1pt}\discretionary{.}{%
}{.}\hspace{.4pt}00004}}


\bibitem{steed2016wild}
A.~Steed, S.~Frlston, M.~M. Lopez, J.~Drummond, Y.~Pan, and D.~Swapp.
\newblock An `in the wild' experiment on presence and embodiment using consumer
  virtual reality equipment.
\newblock {\em IEEE transactions on visualization and computer graphics},
  22(4):1406--1414, 2016.

\bibitem{tai2012daylighting}
N.-C. Tai.
\newblock Daylighting and its impact on depth perception in a daylit space.
\newblock {\em Journal of Light \& Visual Environment}, 36(1):16--22, 2012.

\bibitem{thomas2002surface}
G.~Thomas, J.~H. Goldberg, D.~J. Cannon, and S.~L. Hillis.
\newblock Surface textures improve the robustness of stereoscopic depth cues.
\newblock {\em Human factors}, 44(1):157--170, 2002.

\bibitem{toye1986effect}
R.~C. Toye.
\newblock The effect of viewing position on the perceived layout of space.
\newblock {\em Perception \& Psychophysics}, 40(2):85--92, 1986.

\bibitem{uploadvr2017report}
{UploadVR}.
\newblock {Report: Vive Users Are 95 Percent Male And Spend 5 Hours Per Week in
  VR}.
\newblock
  \href{https://uploadvr.com/vive-users-94-9-percent-male-spend-5-hours-week-vr-average/}{https://uploadvr.com/vive-users-94-9-percent-male-spend-5-hours-week-vr}\href{https://uploadvr.com/vive-users-94-9-percent-male-spend-5-hours-week-vr-average/}{-average/},
  2017.

\bibitem{van2017gamification}
N.~{van Berkel}, J.~Goncalves, S.~Hosio, and V.~Kostakos.
\newblock Gamification of mobile experience sampling improves data quality and
  quantity.
\newblock {\em Proceedings of the ACM on Interactive, Mobile, Wearable and
  Ubiquitous Technologies}, 1(3):1--21, 2017.

\bibitem{wagner1985metric}
M.~Wagner.
\newblock The metric of visual space.
\newblock {\em Perception \& psychophysics}, 38(6):483--495, 1985.

\bibitem{wagner2006geometries}
M.~Wagner.
\newblock {\em The geometries of visual space}.
\newblock Psychology Press, 2006.

\bibitem{wagner2012sensory}
M.~Wagner.
\newblock Sensory and cognitive explanations for a century of size constancy
  research.
\newblock {\em Visual Experience: Sensation, Cognition, and Constancy}, 07
  2012. doi: {{%
10\hspace{.1pt}\discretionary{.}{%
}{.}\hspace{.4pt}1093\discretionary{/}{%
}{/}acprof\discretionary{:}{%
}{:}oso\discretionary{/}{%
}{/}9780199597277\hspace{.1pt}\discretionary{.}{%
}{.}\hspace{.4pt}003\hspace{.1pt}\discretionary{.}{%
}{.}\hspace{.4pt}0004}}


\bibitem{willemsen2008effects}
P.~Willemsen, A.~A. Gooch, W.~B. Thompson, and S.~H. Creem-Regehr.
\newblock Effects of stereo viewing conditions on distance perception in
  virtual environments.
\newblock {\em Presence: Teleoperators and Virtual Environments},
  17(1):91--101, 2008.

\bibitem{woods2015conducting}
A.~T. Woods, C.~Velasco, C.~A. Levitan, X.~Wan, and C.~Spence.
\newblock Conducting perception research over the internet: a tutorial review.
\newblock {\em PeerJ}, 3:e1058, 2015.

\end{thebibliography}
